\documentclass{aa}

\usepackage{natbib}
\bibpunct{(}{)}{;}{a}{}{,} 
\usepackage{graphics}   
\usepackage{graphicx}   
\usepackage{epstopdf}
\usepackage{longtable}   
\usepackage{url}      
\usepackage{bm}        
\usepackage[varg]{txfonts}
\usepackage{pdflscape}
\usepackage{supertabular}
\usepackage{hyperref}   
\usepackage{amsmath}
\usepackage[svgnames]{xcolor}

\hypersetup{
    bookmarks=true,         
    unicode=false,          
    colorlinks=true,       
    linkcolor=blue,          
    citecolor=blue,        
    filecolor=blue,      
    urlcolor=blue           
}
\usepackage{color}

\begin{document}

\title{Grids of stellar models with rotation} 
\subtitle{VI. Models from $0.8$ to $120\,M_\odot$ at a metallicity $Z = 0.006$}

\author{
Patrick Eggenberger\inst{1}, Sylvia Ekstr{\"o}m\inst{1}, Cyril Georgy\inst{1}, S\'ebastien Martinet\inst{1}, Camilla Pezzotti\inst{1}, Devesh Nandal\inst{1}, Georges Meynet\inst{1}, Ga{\"e}l Buldgen\inst{1}, S\'ebastien Salmon\inst{1}, Lionel Haemmerl\'e\inst{1}, Andr{\'e} Maeder\inst{1}, Raphael Hirschi\inst{2, 3}, Norhasliza Yusof\inst{4}, Jos{\'e} Groh\inst{5}, Eoin Farrell\inst{5}, Laura Murphy\inst{5}, Arthur Choplin\inst{6}}

\institute{Department of Astronomy, University of Geneva, Chemin Pegasi 51, CH-1290 Versoix, Switzerland
\and
Astrophysics Group, Keele University, Keele, Staffordshire, ST5 5BG, UK
\and
Institute for the Physics and Mathematics of the Universe (WPI), University of Tokyo, 5-1-5 Kashiwanoha, Kashiwa 277-8583, Japan
\and
Department of Physics, Faculty of Science, University of Malaya, 50603 Kuala Lumpur, Malaysia
\and
School of Physics, Trinity College Dublin, the University of Dublin, College Green, Dublin
\and
Institut d'Astronomie et d'Astrophysique, Universit\'e Libre de Bruxelles (ULB), CP 226, 1050 Brussels, Belgium}

\titlerunning{Grids of stellar models at Z=0.006}
\authorrunning{Eggenberger et al.}

\date{Received /Accepted}

\abstract{Grids of stellar models, computed with the same physical ingredients, allow one to study
 the impact of a given physics on a broad range of initial conditions and they are a key ingredient for modeling the evolution of galaxies.}
{
We present here a grid of single star models for masses between 0.8 and 120 M$_\odot$, with and without rotation for a mass fraction of heavy element $Z$=0.006, representative of the Large Magellanic Cloud (LMC).}
{
We used the GENeva stellar Evolution Code (GENEC). The evolution was computed until the end of the central carbon-burning phase, the early asymptotic giant branch phase, or the core helium-flash for massive, intermediate, and low mass stars, respectively.}
{The outputs of the present stellar models are well framed by the outputs of the two grids obtained by our group for metallicities above and below the one considered here.
The models of the present work provide a good fit to the nitrogen surface enrichments observed during the main sequence for stars in the LMC with initial masses around 15 M$_\odot$. They also reproduce the slope of the luminosity function of red supergiants of the LMC well, which is a feature that is sensitive to the time-averaged mass loss rate over the red supergiant phase. The most massive black hole that can be formed from the present models at $Z$=0.006 is around 55 M$_\odot$. No model in the range of mass considered will enter into the pair-instability supernova regime, while the minimal mass to enter the region of pair pulsation instability is around 60 M$_\odot$ for the rotating models and 85 M$_\odot$  for the nonrotating ones.
}
{The present models are of particular interest for comparisons with observations in the LMC and also in the outer regions of the Milky Way. We provide public access to numerical tables that can be used for computing interpolated tracks and for population synthesis studies.
}

\maketitle

\section{Introduction}

Grids of single stellar models \citep[e.g.][]{Heger2003, Brott2011, Limongi2006, Bressan2012, Limongi2012, Chieffi2013, Choi2016, Limongi2018, Hidalgo2018, Claret2019,Szecsi2020}, covering large domains of initial masses, metallicities, and rotations, and those computed with exactly the same physics and physical ingredients provide a global view of the consequences of a given physics in both low and high mass stars, in metal-poor as well as in metal-rich regions, and in different phases of evolution. Their outputs can then be compared with many different observed features providing insights in the dependence on metallicity of the evolution of stars. Large stellar grids at different metallicities are also important ingredients of population synthesis models \citep[e.g.][and references therein]{Bruzual2003,Voss2009, Leitherer2012, Georgy2014, Pacifici2015, Vazquez2017, Leitherer2019}, which are essential to predict the radiative, mechanical, and chemical feedback of stars in galaxies during the whole cosmic history and to associate an age to systems going from planetary systems to starbursts.

The present grid complements the grids already published for metallicities $Z$=0.014  \citep[papers I and II,][]{Ekstrom2012a, Georgy2012}, $Z=0.002$ \citep[paper III,][]{Georgy2013a}, $Z=0.0004$ \citep[paper IV,][]{groh2019}, and $Z=0$ \citep[paper V,][]{Murphy2020}. The physical ingredients used are identical in all these grids, which only differ by the initial chemical composition adopted. This paper presents stellar evolutionary tracks for a chemical composition characterized by a mass fraction of heavy elements $Z$=0.006, that is for a metallicity that is near one of the stars in the Large Magellanic Cloud (LMC).
The models are available under the form of electronic tables and can be used as input for interpolating tracks for any given initial masses between 0.8 and 120 M$_\odot$, and for computing isochrones and synthetic color magnitude diagrams of open clusters using the SYCLIST code \citep{Georgy2014}. These grids and codes can be used through a web-interface\footnote{https://www.unige.ch/sciences/astro/evolution/en/database/}.

The paper is organized in the following way: Section~\ref{sec:physics} briefly recalls the main physical ingredients used. The global features of the stellar models are discussed in Sect.~\ref{sec:z0p006}, while a comparison between the $Z=0.006$ models and the $Z=0.002$ and 0.014 grids is provided in Sect.~\ref{sec:metallicity}. Section~\ref{sec:obs} compares the outputs of the present grid to the surface abundances of main-sequence (MS) stars in the LMC and to the red supergiant luminosity function in the LMC. The main conclusions are given in Sect.~\ref{sec:conclusion}.

\section{Ingredients of the stellar models}
\label{sec:physics}

The models have been computed with the same physical ingredients as in \cite{Ekstrom2012a}, but for initial abundances 
$X= 0.738$, $Y=0.256$ and $Z=0.006$. The heavy elements distribution is solar scaled with the solar abundances from \cite{asp05} except for the neon which is taken from \cite{cun06}. A fixed value of the overshooting parameter has been adopted to extend the size of the convective cores during the H- and He-burning phases by an amount equal to 10\% the pressure scale height estimated at the Schwarzschild boundary. Such an assumption of a constant overshooting parameter for the entire mass domain is of course questionable, with studies favoring an overshooting that increases with the stellar mass above about 9 M$_\odot$ \citep[][]{Brott2011, Castro2014,mar21,sco21}. The present grid does not consider such a variation in the overshooting parameter for the sake of homogeneity with the physics adopted in the first paper of this series \citep{Ekstrom2012a}. Keeping a constant physics is indeed required to demonstrate the effects of changing the mass, the initial composition or the initial rotation rate. 

The transport of chemical species and angular momentum, as well as the impact of rotation on the mass
loss rates have also been taken as in \citet{Ekstrom2012a}. We recall that the present rotating models only account for hydrodynamical transport by the shear instability and meridional currents, and do not take into account the possible impact of internal magnetic fields. Consequently, the present models exhibit a significant degree of radial differential rotation in their radiative interior during the MS phase in contrast to models computed in the framework of the Tayler-Spruit dynamo \citep{Spruit2002} for instance, which are characterized by an almost uniform internal rotation during the main part of the evolution on the MS.

\begin{figure}
\includegraphics[width=0.45\textwidth]{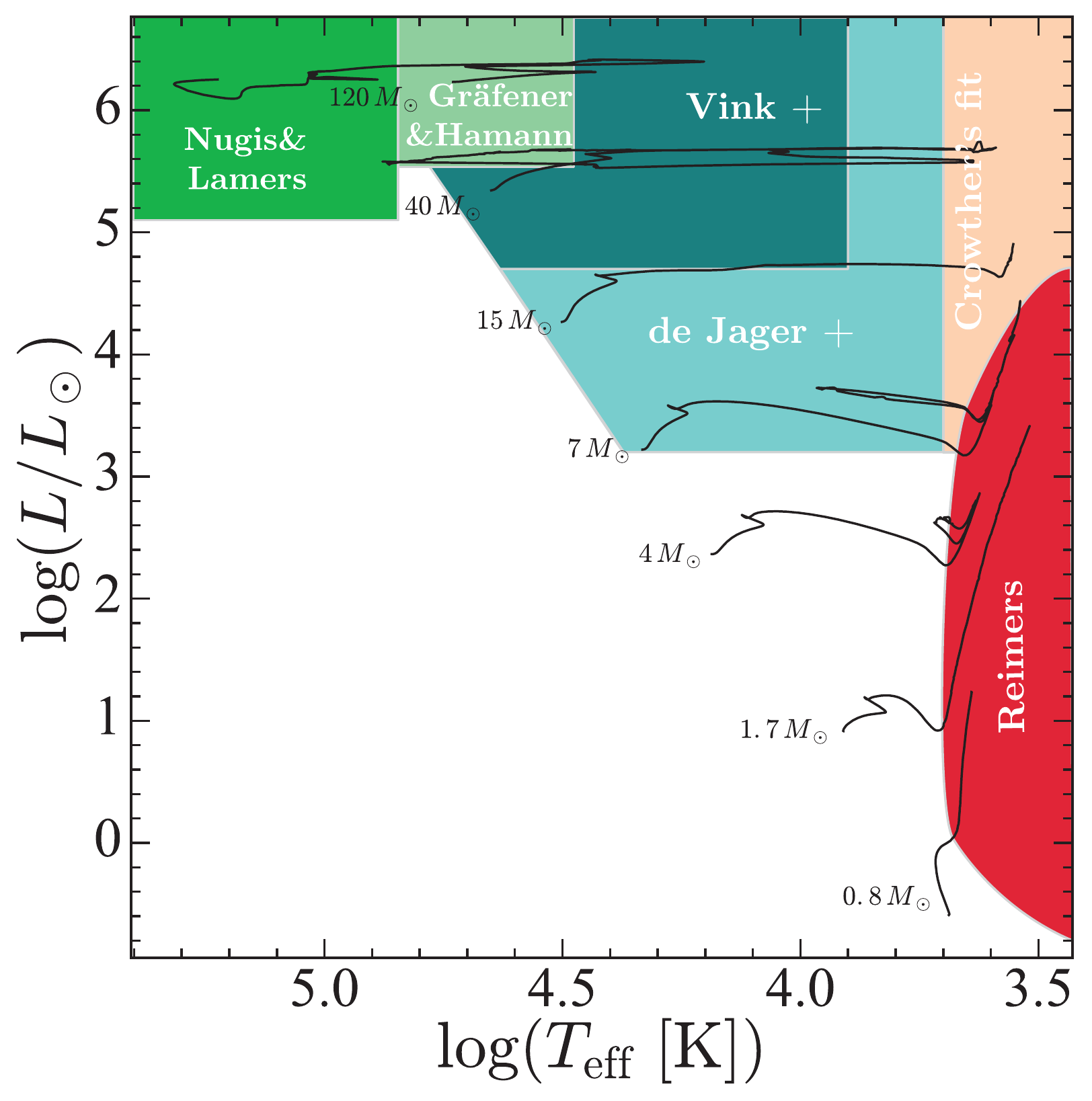}
\caption{Domains of application of the various mass loss rates prescriptions in the HR diagram. The mass loss rates are from
\citet{Reimers1975, deJager1988, Nugis2000, Vink2001,Nugis2002, cro01, Grafener2008}.}
\label{masslossdomains}
\end{figure}

The mass loss rates have been taken as indicated in Fig.~\ref{masslossdomains} \citep[see also][]{Ekstrom2012a}.
The dependence on metallicity has been taken so that 
$
\dot{M}(Z)=(Z/Z_\odot)^{0.7}\dot{M}(Z_\odot)
$ except during the red giant and supergiant phase for which no dependence on the metallicity has been considered. According to \citet{ vanLoon2005} and \citet{ Groen2012, Groen2012cor}, the metallicity dependence for these stars do indeed appear to be weak.

\section{The $Z$=0.006 grid of stellar models}
\label{sec:z0p006}

\begin{figure*}
\includegraphics[width=0.48\textwidth]{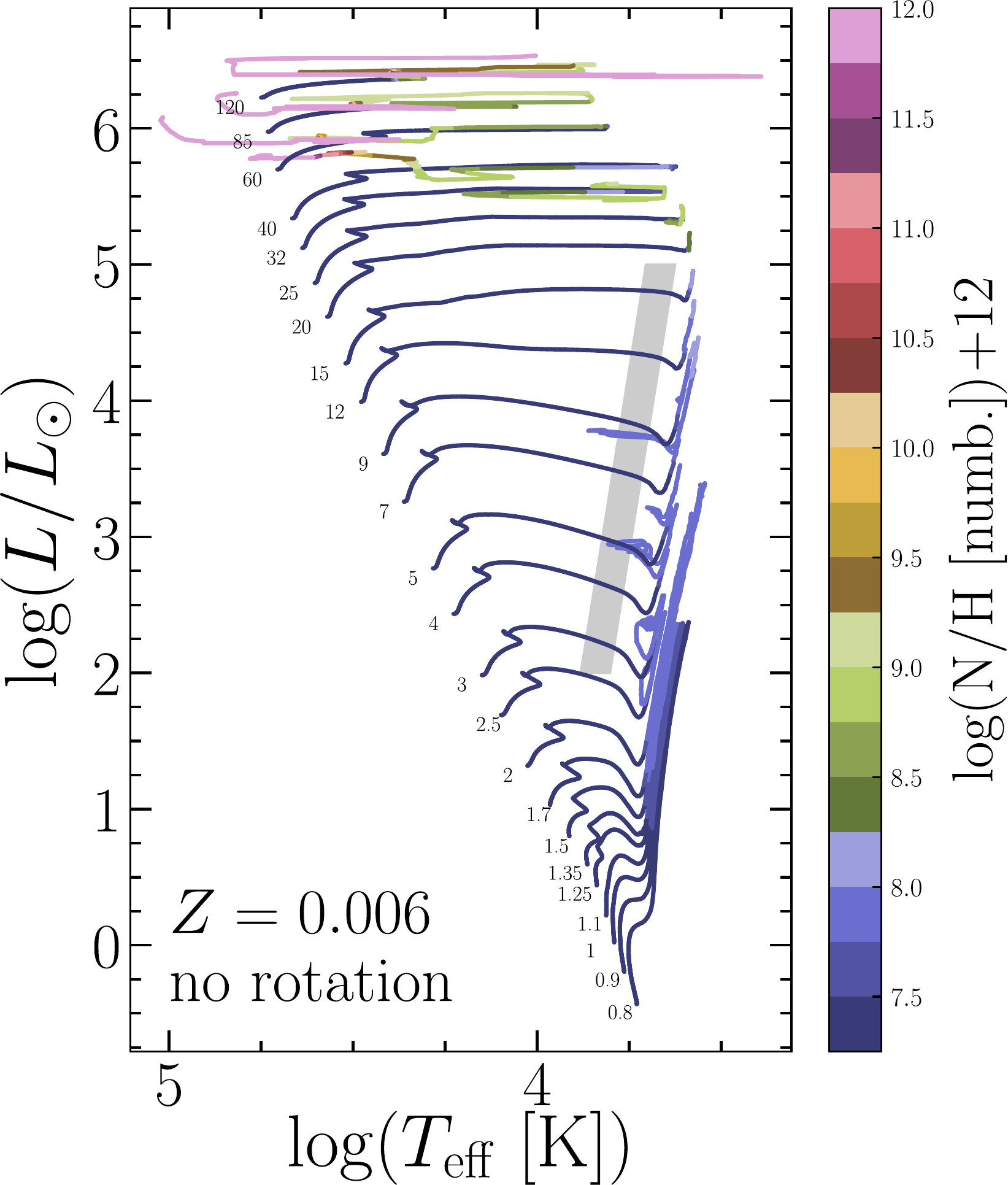}\hfill\includegraphics[width=0.48\textwidth]{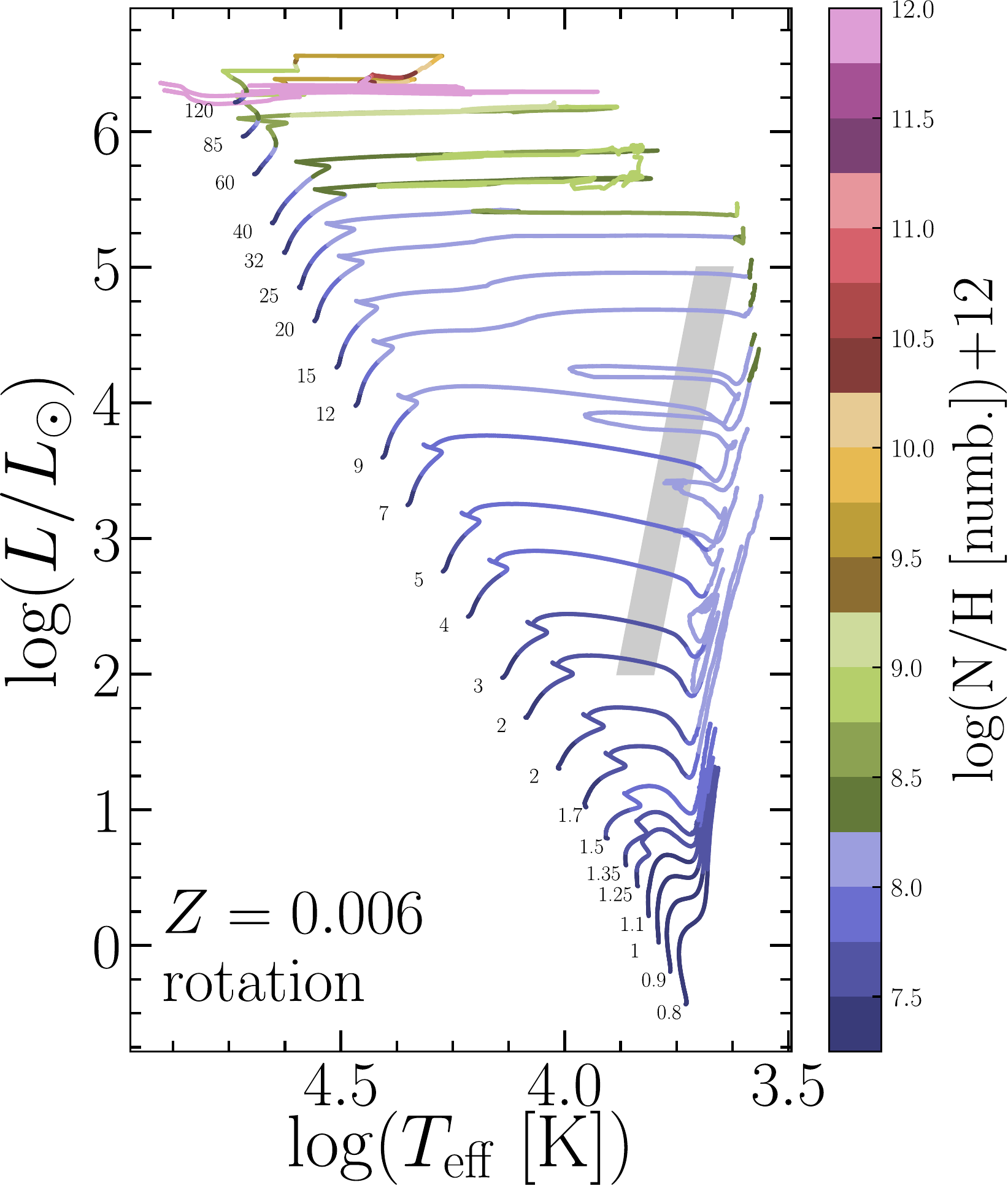}
\caption{Evolutionary tracks for models with an initial mass between 0.8 and 120 M$_\odot$ with a mass fraction of heavy element $Z$=0.006. The colors along the tracks give
an indication of the surface nitrogen to hydrogen ratios. The gray band shows the Cepheid instability strip.
\textsl{Left panel:}  Nonrotating models.
\textsl{Right panel:} Rotating models with $\upsilon_{\rm ini}/\upsilon_{\rm crit}$=0.4. The initial equatorial velocity on the ZAMS, as well as the time averaged velocity on the MS is given in the second and third column of Table~\ref{TabListModels}.} 
\label{Fig:HRDgen}
\end{figure*}

\begin{figure}
\includegraphics[width=0.24\textwidth]{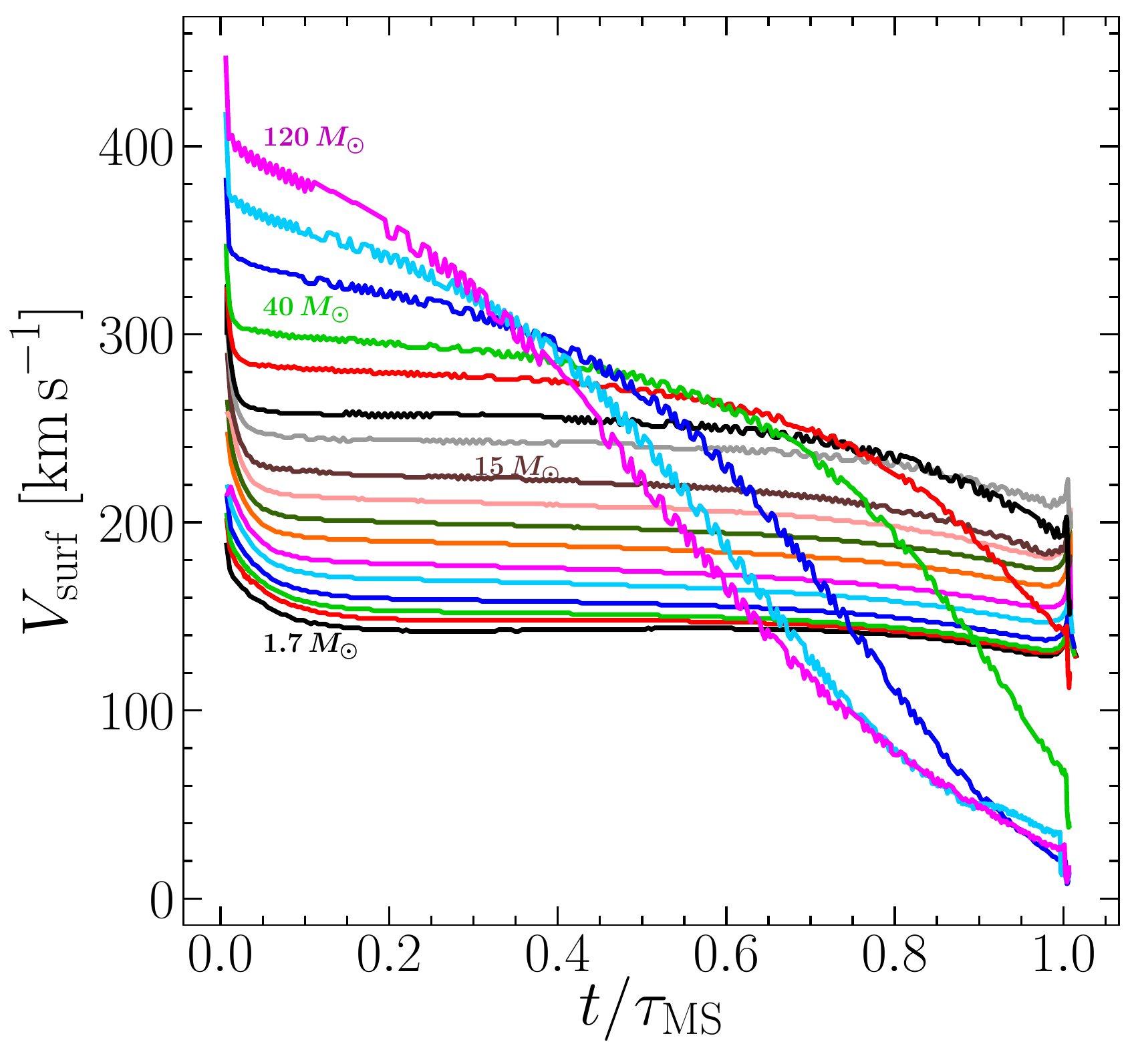}\includegraphics[width=0.24\textwidth]{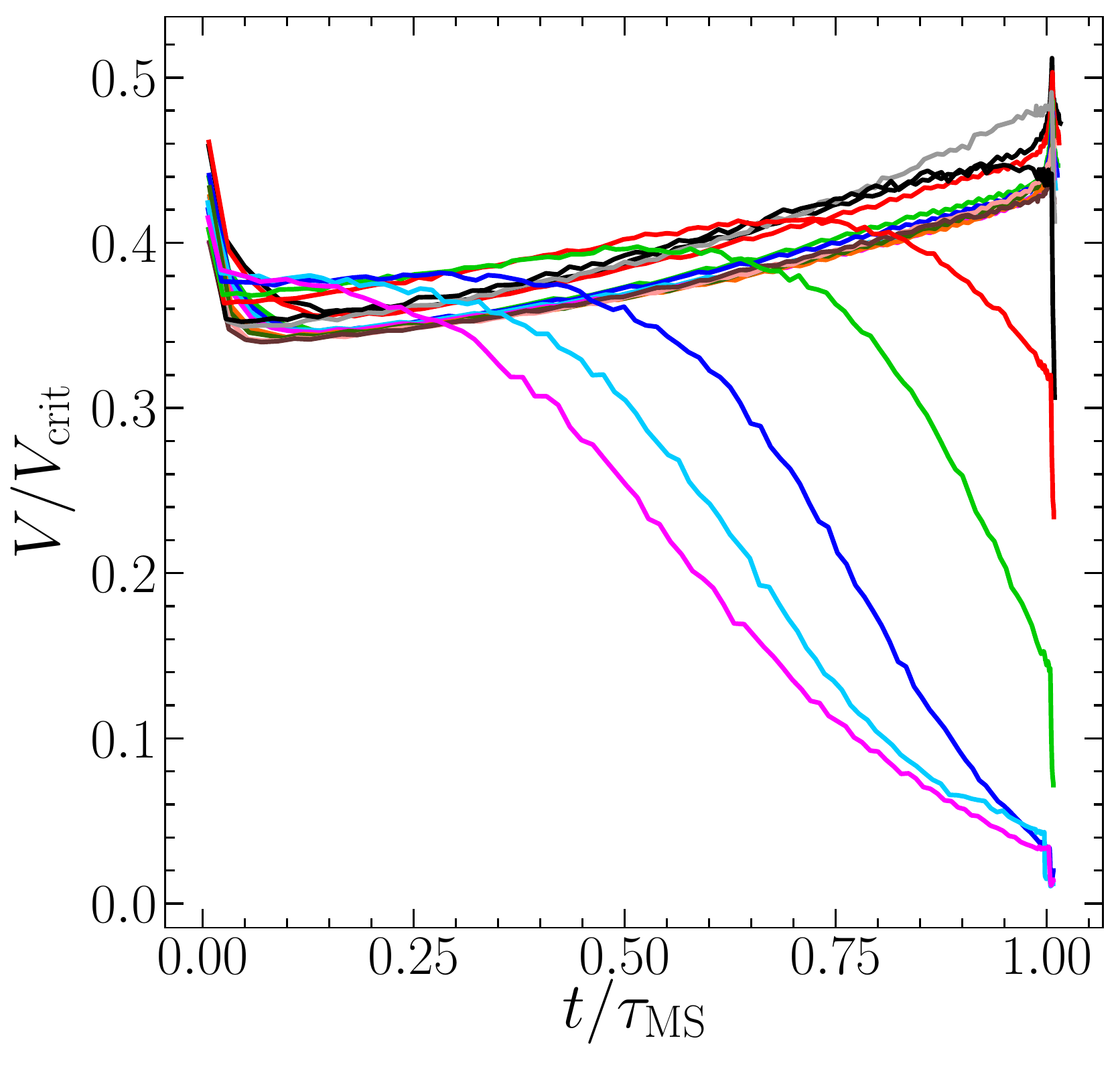}
\caption{\textsl{Left panel:} Evolution of the surface equatorial velocity as a function of time (normalized to the MS lifetime for each mass) for rotating models of different masses at $Z$=0.006. \textsl{Right panel:} Same as left panel but for the ratio of the surface equatorial velocity to the critical velocity. 
}
\label{Fig:vsurf}
\end{figure}

A few characteristics of the models on the zero-age main sequence (ZAMS) and at the end of the core H-, He- and C-burning phases are presented in Table~\ref{TabListModels}. The tracks in the HR diagram (HRD) are shown
in Fig.~\ref{Fig:HRDgen}. 
The comparison of nonrotating and rotating models with $\upsilon_{\rm ini}/\upsilon_{\rm crit}$=0.4 provide some striking differences. Firstly, as is well known, the tracks of rotating models show sign of surface nitrogen enrichments on a much broader mass domain than nonrotating tracks. Let us for instance look at the least massive model
for which the N/C is more than twice the initial one at the end of the MS phase according to Table~\ref{TabListModels}. The least massive model in the nonrotating case is between 60 and 85 M$_\odot$, while in the rotating case, it is between 1.35 and 1.5 M$_\odot$. This is of course a direct consequence of rotational mixing.
Secondly, in the upper mass range (above or equal to 40 M$_\odot$), the nonrotating tracks appear to show surprisingly stronger surface  enrichments than rotating ones. 
At first sight, we would expect that rotation, favoring internal mixing, would also favor stronger surface enrichments. For these massive stars, the mass losses by stellar winds uncover deeper layers whose composition has been changed either by mixing and/or nuclear reactions. In the present case, mass loss plays the dominant role and the rotating models lose less mass than nonrotating ones. This may appear counter-intuitive, because, at a given position in the HRD, rotation increases the quantity of mass lost by stellar winds. Thus, in case where rotation would not affect the shape of the tracks in the HRD, rotating models would always lose more mass than nonrotating ones. However, the MS tracks of rotating models with masses above about 40 M$_\odot$ are bluer than the nonrotating ones because of rotational mixing. At a given luminosity, a shift to the blue part of the HRD results in a decrease in the mass loss rates, so that rotating models lose less mass and thus show weaker surface nitrogen enrichments than nonrotating models. Finally, cepheid loops are more extended in the grid of rotating stellar models. This is an effect that is not present (or at least not in such an obvious way) in the grids at $Z=0.002$ and $0.014$. These blue loops are very sensitive to details in the physics of the models \citep[e.g.][]{wal15} and thus it is not a real surprise that these loops can be differently affected by rotation at various metallicities.

Figure~\ref{Fig:vsurf} (left panel) shows the evolution of the surface equatorial velocity as a function of time for different initial mass models. Increasing the initial mass also increases the initial surface velocity. This comes from the assumption of a constant ratio between the surface rotation and the critical rotation velocity at the ZAMS. Since the critical velocity at the ZAMS increases with the initial mass, higher initial rotation velocities are obtained for more massive stars.

For masses between 1.7 and 20 M$_\odot$, the surface velocity remains more or less constant during the main part of the MS evolution. Just at the very beginning of the MS evolution, there is a rapid drop of the surface velocity due to a redistribution of angular momentum inside the star by meridional currents \citep{den99}. After this short phase, since the radius increases as the star evolves on the MS, local conservation of angular momentum implies a significant decrease in the surface velocity. This decrease is however counterbalanced by the transport of angular momentum from the core to the envelope mainly by the meridional currents. Since for these masses, stellar winds remain modest, not much angular momentum is removed from the star and thus the surface velocity decreases only very slowly.

For stars with masses above about 40 M$_\odot$, the surface velocity exhibits a strong decrease during the MS phase. This is an effect of the mass lost by stellar winds that becomes more and more important as the initial mass of the star increases.  As a numerical example, a 40 M$_\odot$ star, observed at the end of the MS phase with a surface velocity of 100 km s$^{-1}$, may have begun its MS evolution with a surface velocity around 300 km s$^{-1}$.

Figure~\ref{Fig:vsurf} (right panel) shows the MS evolution of the ratio of the surface equatorial velocity to the critical velocity. For initial masses that suffer little mass losses, this ratio increases as a function of time. In those cases, the decrease in the critical velocity when the star inflates is more rapid than the slow decrease in the surface velocity. For the more massive stars, as indicated above, mass losses are  much larger and the ratio of surface to critical velocity decreases as a function of time.

\section{Effects of metallicity}
\label{sec:metallicity}

\subsection{Effects of metallicity versus rotation in the HR diagram}

\begin{figure}
\begin{center}
\includegraphics[width=0.24\textwidth]{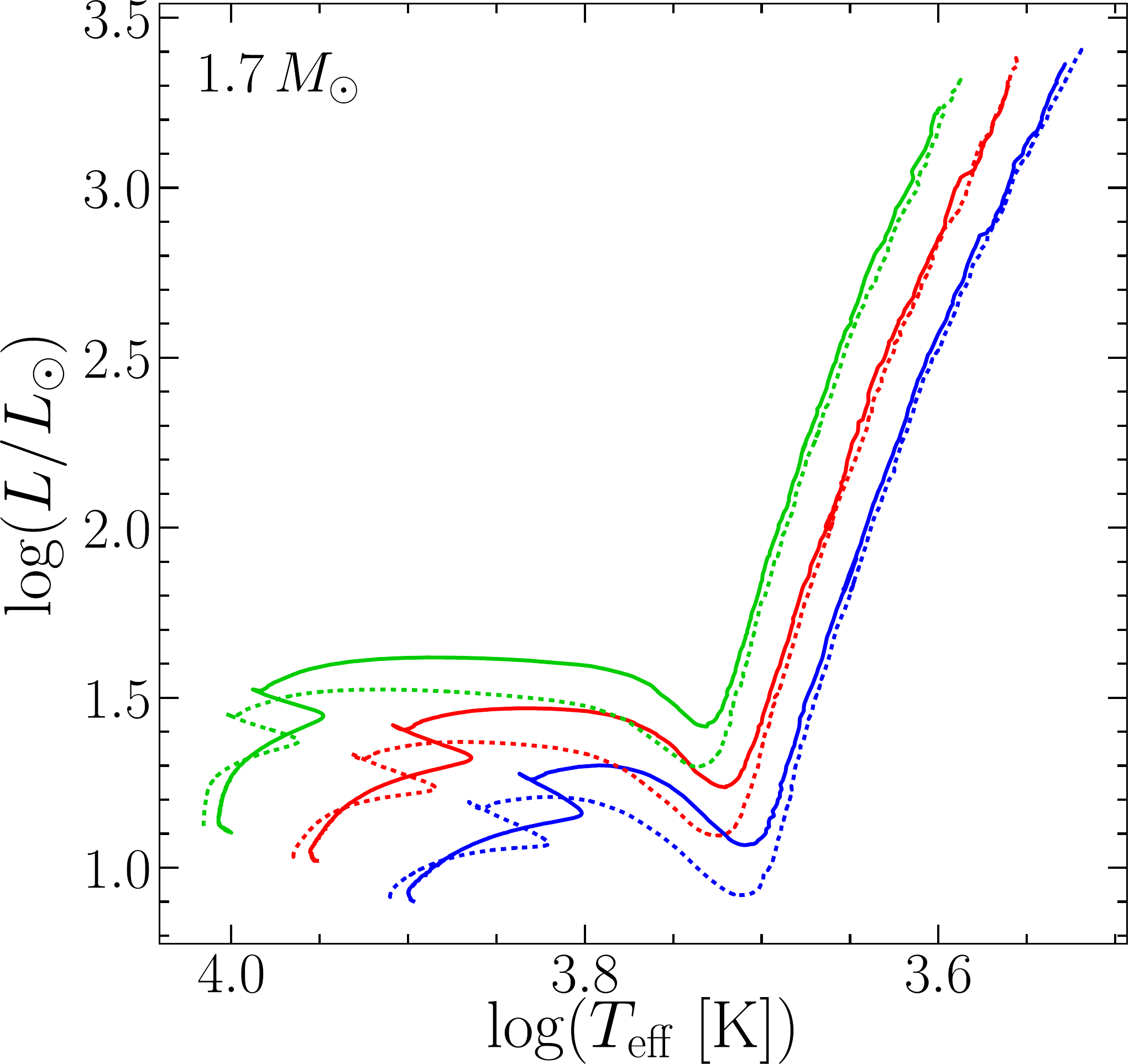}\includegraphics[width=0.24\textwidth]{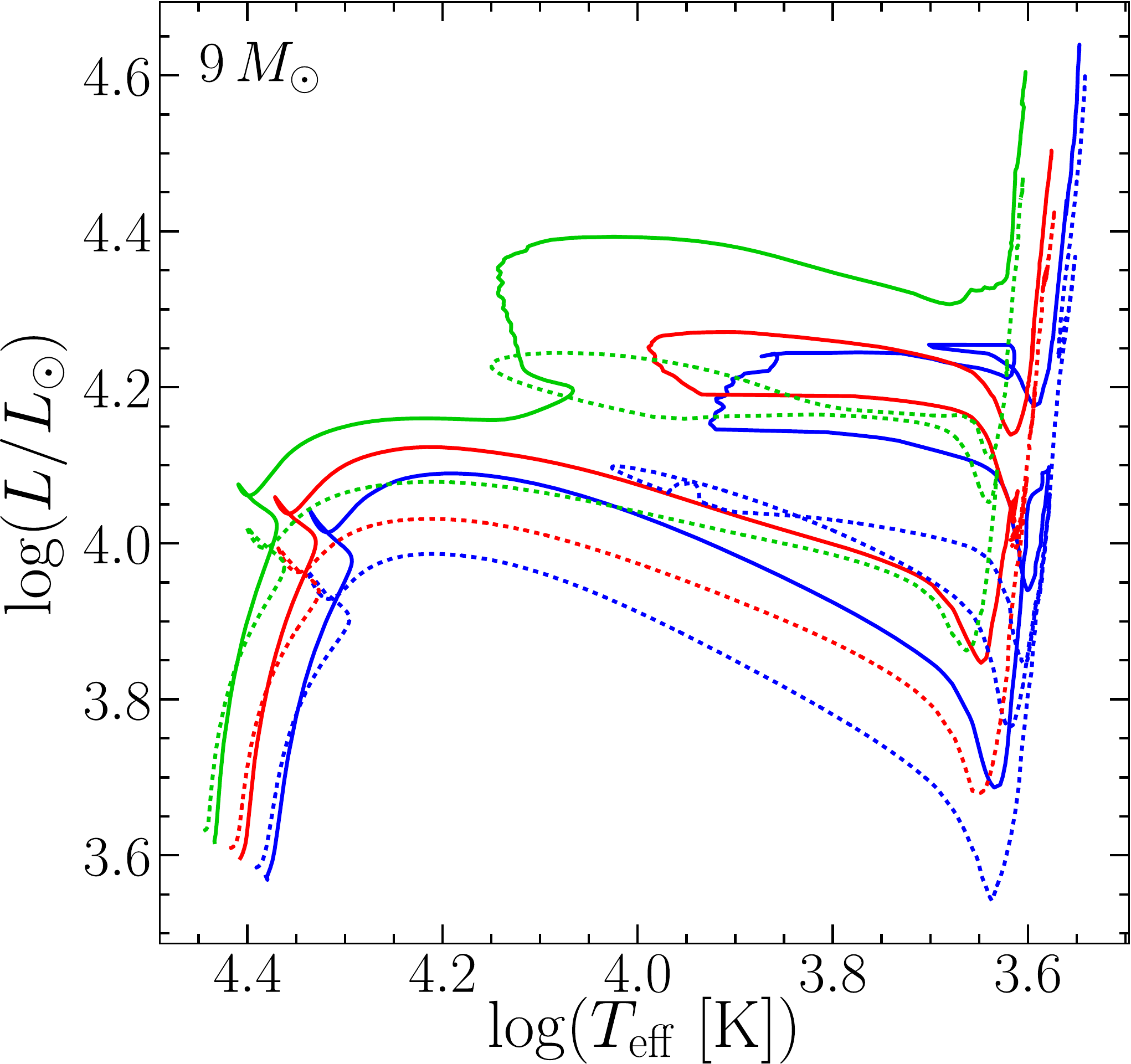}
\includegraphics[width=0.24\textwidth]{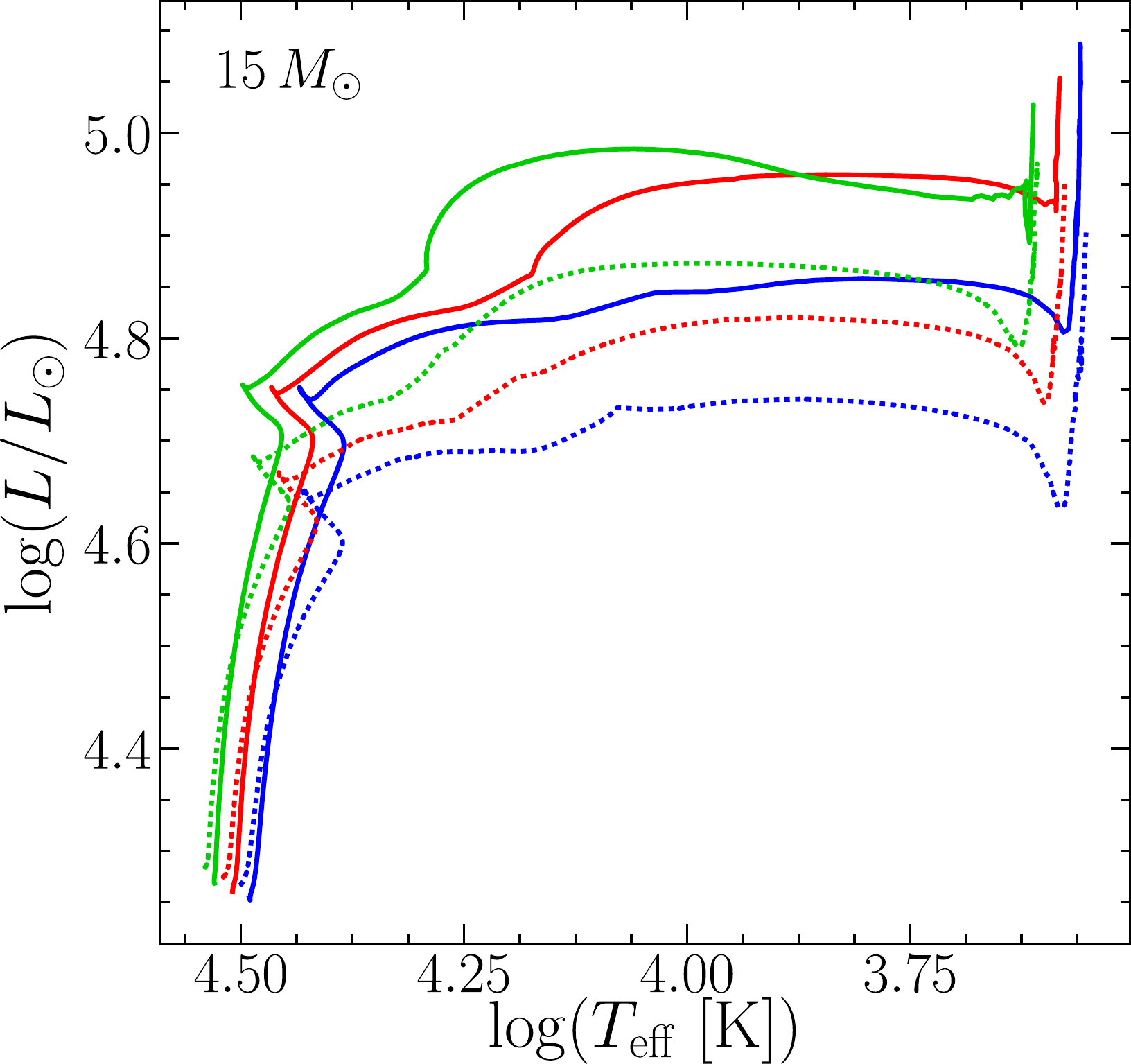}\includegraphics[width=0.24\textwidth]{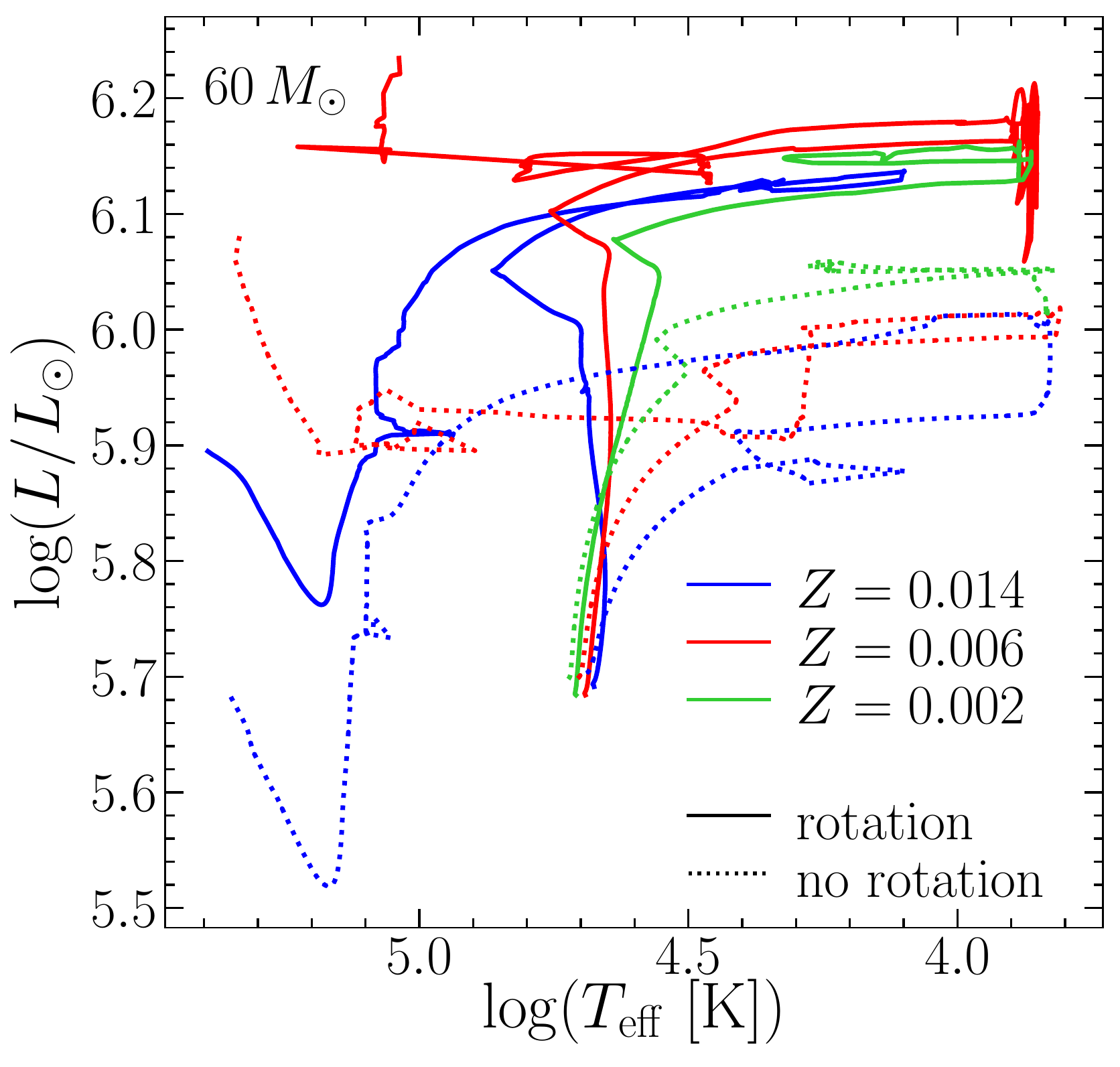}
\caption{Comparisons between evolutionary tracks in the HRD at $Z=0.014$ (models of paper I, blue lines), $Z=0.006$ (models discussed in this paper, red lines) and $Z=0.002$ (models of paper III, green lines) with (solid lines) and without rotation (dotted lines) for 1.7, 9, 15 and 60 M$_\odot$ models.
}
\label{Fig:HRDcomp}
\end{center}
\end{figure}

Figure~\ref{Fig:HRDcomp} compares tracks in the HRD for models with the same initial mass computed with and without rotation at three different metallicities. Decreasing the metallicity shifts the tracks to hotter and more luminous parts of the HRD during the MS phase. In general, during the MS, a change in metallicity (between the three values of $Z$ considered here) has a larger impact on the tracks than the inclusion of rotation. Only for the most massive models (illustrated in Fig.~\ref{Fig:HRDcomp} with the 60 M$_\odot$ models) rotation can have a larger effect on the tracks than a change in metallicity.

After the MS, the effects of rotation become important, at least as important as changing the metallicity. 
For the 9 M$_\odot$ models, rotation significantly increases the luminosity after the MS phase.
For instance, a rotating 9 M$_\odot$ with $Z$=0.014 at an effective temperature Log $T_{\rm eff}$ equal to 4.0 during the first crossing of the HR gap can be as luminous as
a nonrotating 9 M$_\odot$ at a lower metallicity of $Z$=0.002. This of course has a consequence for the properties of Cepheids that are formed in different metallicity environments and with different angular momentum contents \citep[e.g.][]{Anderson2014, Anderson2016, Anderson2020}. The 15 M$_\odot$ models no longer exhibit blue loops after the MS, but show similar increases in the luminosity when the metallicity is decreased or when the effects of rotation are accounted for. It is worthwhile to discuss the case of the 60 M$_\odot$ models in a more detailed way. First, during the MS, as already mentioned above, the impact of rotation largely blurs the effects due to a change in metallicity.
The 60 M$_\odot$ tracks are nearly vertical in the HRD during the MS, which reflects the fact that the stars follow a nearly chemically-homogeneous evolution. Interestingly, the trend toward homogeneous evolution during the MS is the strongest for the $Z$=0.014 tracks and then decreases when the metallicity decreases from $Z=0.014$ to 0.006 and 0.002. This results from the combined effects of rotational mixing and mass loss. The characteristic timescale for rotation induced transport of chemicals is in general shorter when the metallicity decreases, mainly because stars are more compact at low metallicity, which leads to less extended radiative envelopes, while gradients of angular momentum are found to be equally strong or even stronger \cite[see][]{MMVII2001}. Thus, rotational effects tend to favor homogeneous evolution at low metallicities\footnote{Another rotational effect can however favor a stronger chemical mixing at high metallicity: stars are less compact at high metallicity and thus meridional currents are stronger in the outer layers (the velocity of the meridional currents varies as the inverse of the density in these external layers).}. 
In contrast, mass loss by stellar winds is stronger at high metallicity. Stronger mass loss uncovers deep layers more rapidly and favors thereby homogeneous evolution at high metallicity. The 60 M$_\odot$ models correspond to a case where the effects of mass loss dominate over the effects of rotation, resulting in more homogeneous stars at high metallicity by removing the outer layers.

\subsection{Changes in surface chemical composition}

\begin{figure}
\includegraphics[width=0.45\textwidth]{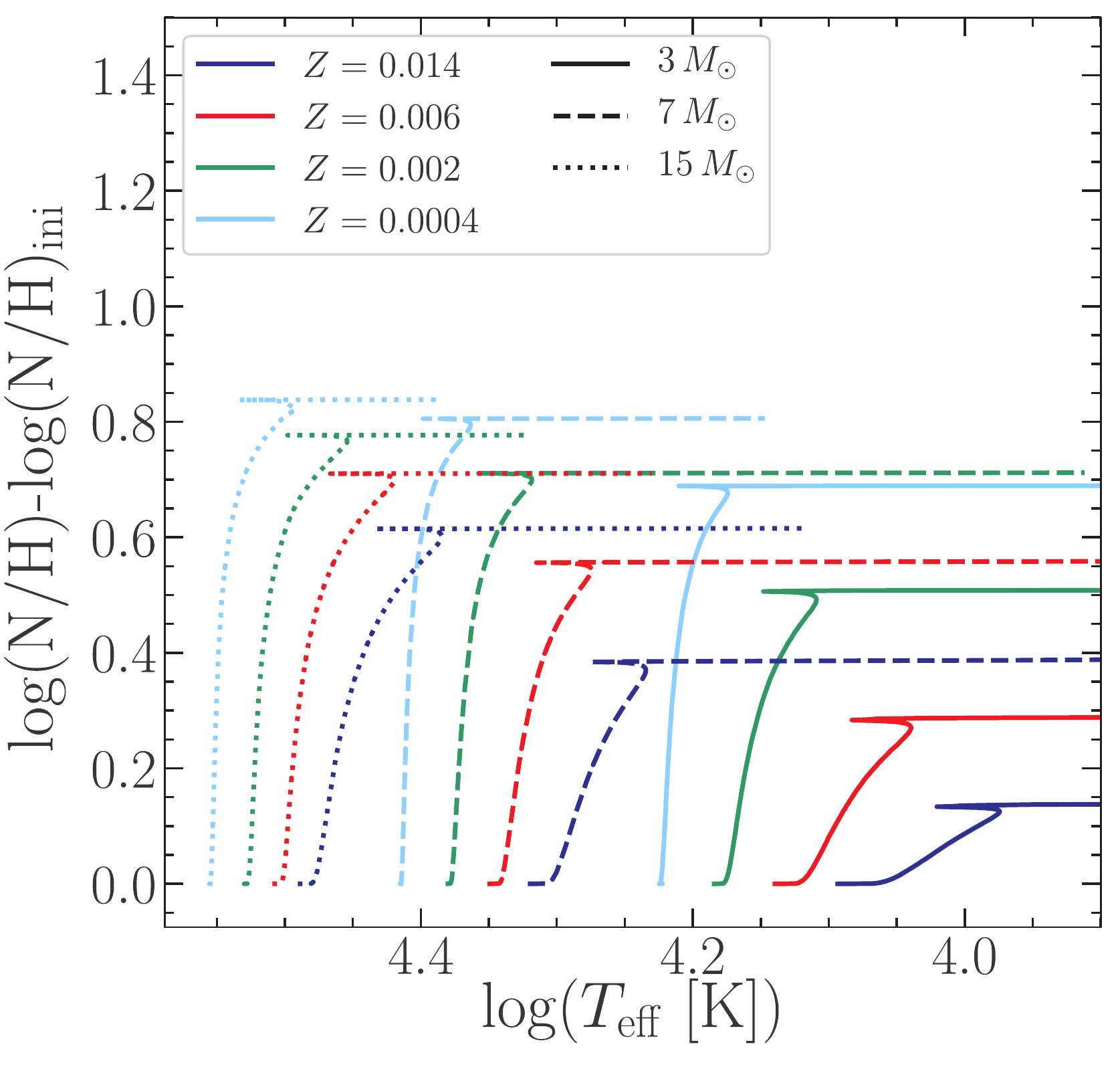}
\caption{Evolution of the N/H ratio as a function of the effective temperature at the surface of rotating models with three different initial masses at four different initial metallicities.} 
\label{Fig:surfab}
\end{figure}

Figure~\ref{Fig:surfab} shows the variation in the surface N/H ratios for rotating stars with initial masses between 3 and 15 M$_\odot$ and $Z$ between 0.0004 and 0.014.
The present $Z=0.006$ models are well framed by the $Z$ equal 0.002 and 0.014 models. As found in previous works, for a given initial rotation velocity, the nitrogen surface enrichment reached at a given age increases when the initial mass increases and the metallicity decreases. Interestingly, the impact of a change in metallicity on the surface N/H ratio is larger for lower mass stars, while the impact of a change in the initial mass on the N/H ratio is larger at higher metallicities.

\subsection{Red and blue supergiants at various metallicities}

\begin{figure*}
\includegraphics[width=0.30\textwidth]{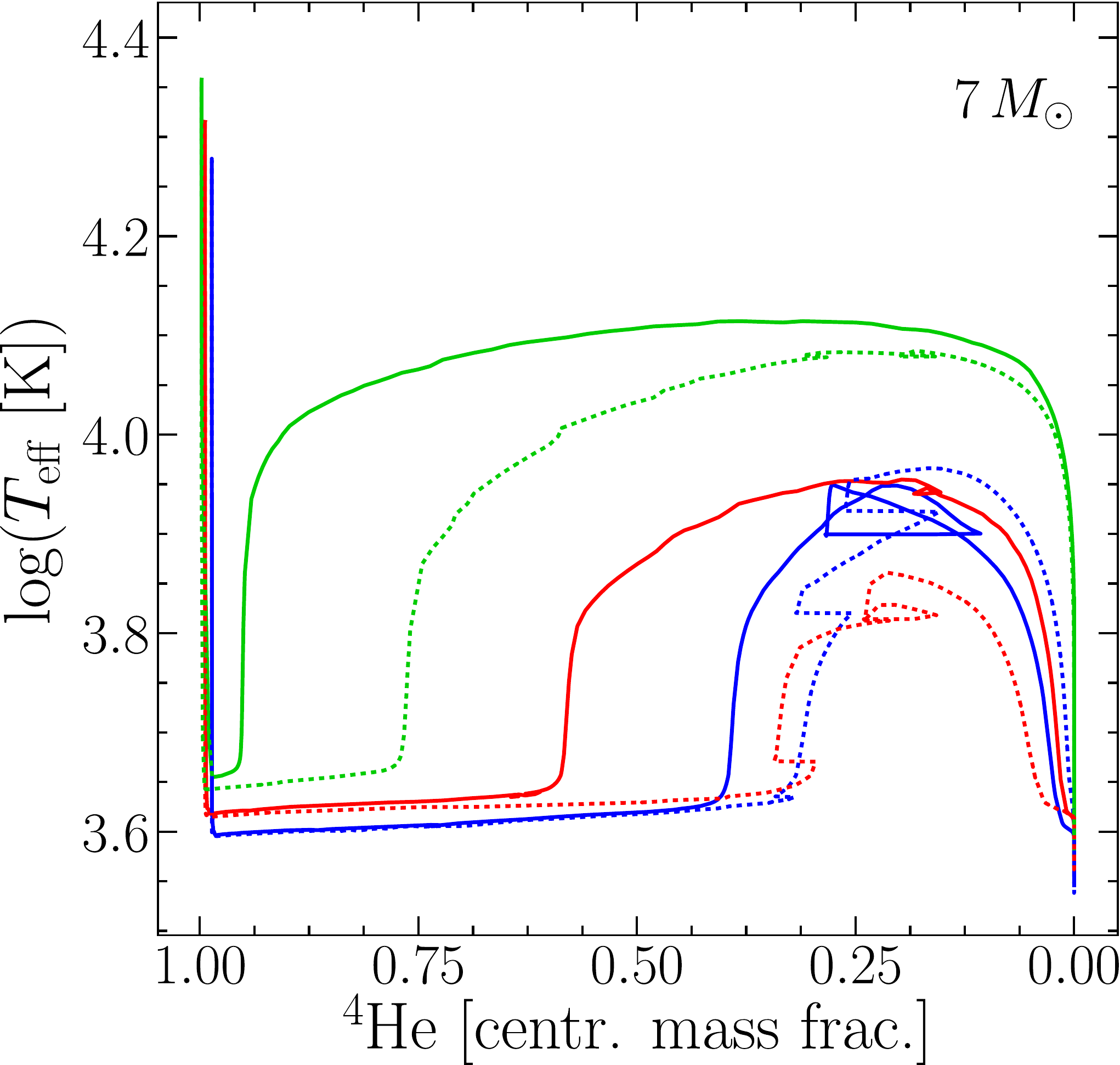}\hfill\includegraphics[width=0.30\textwidth]{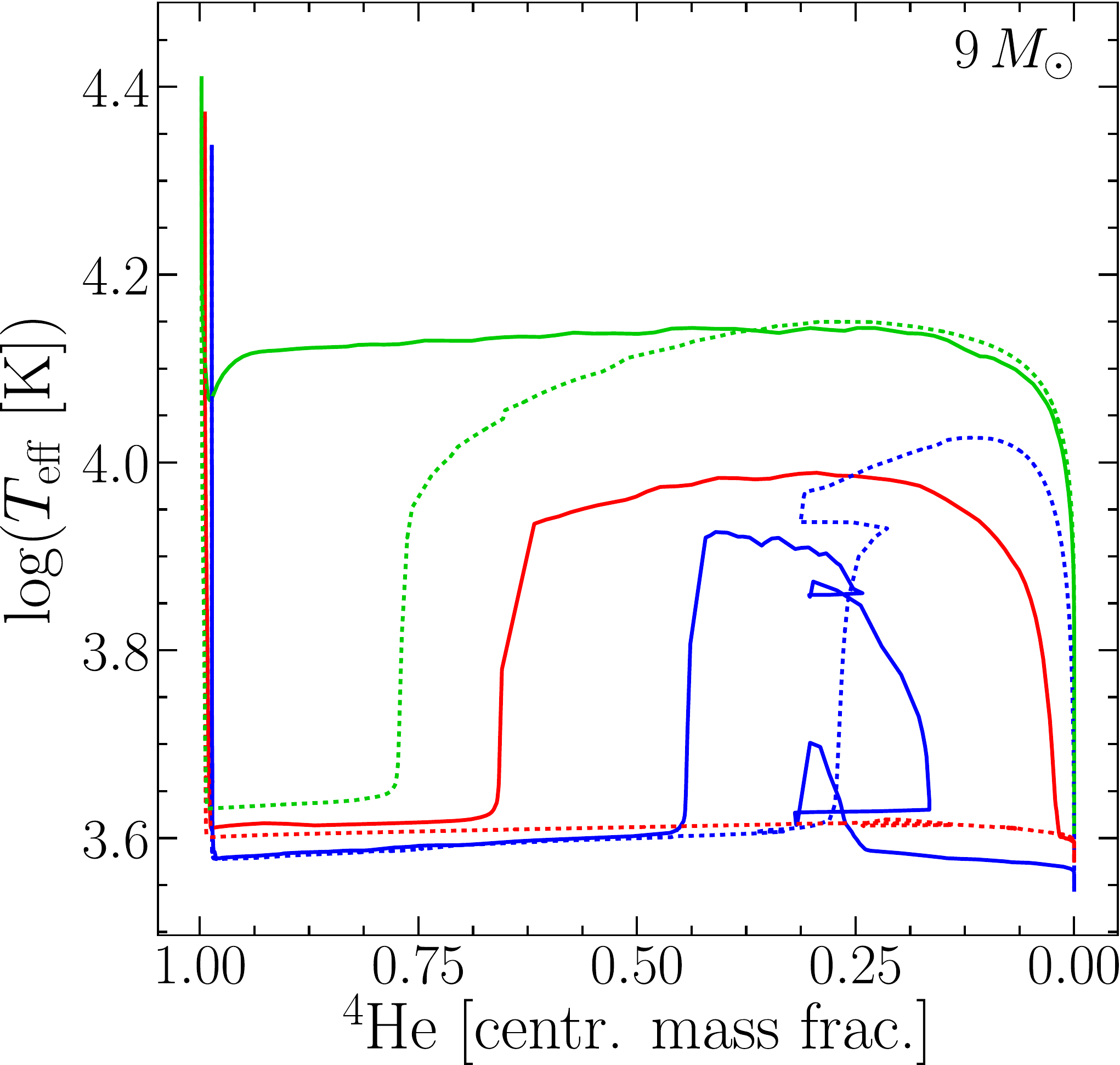}\hfill\includegraphics[width=0.30\textwidth]{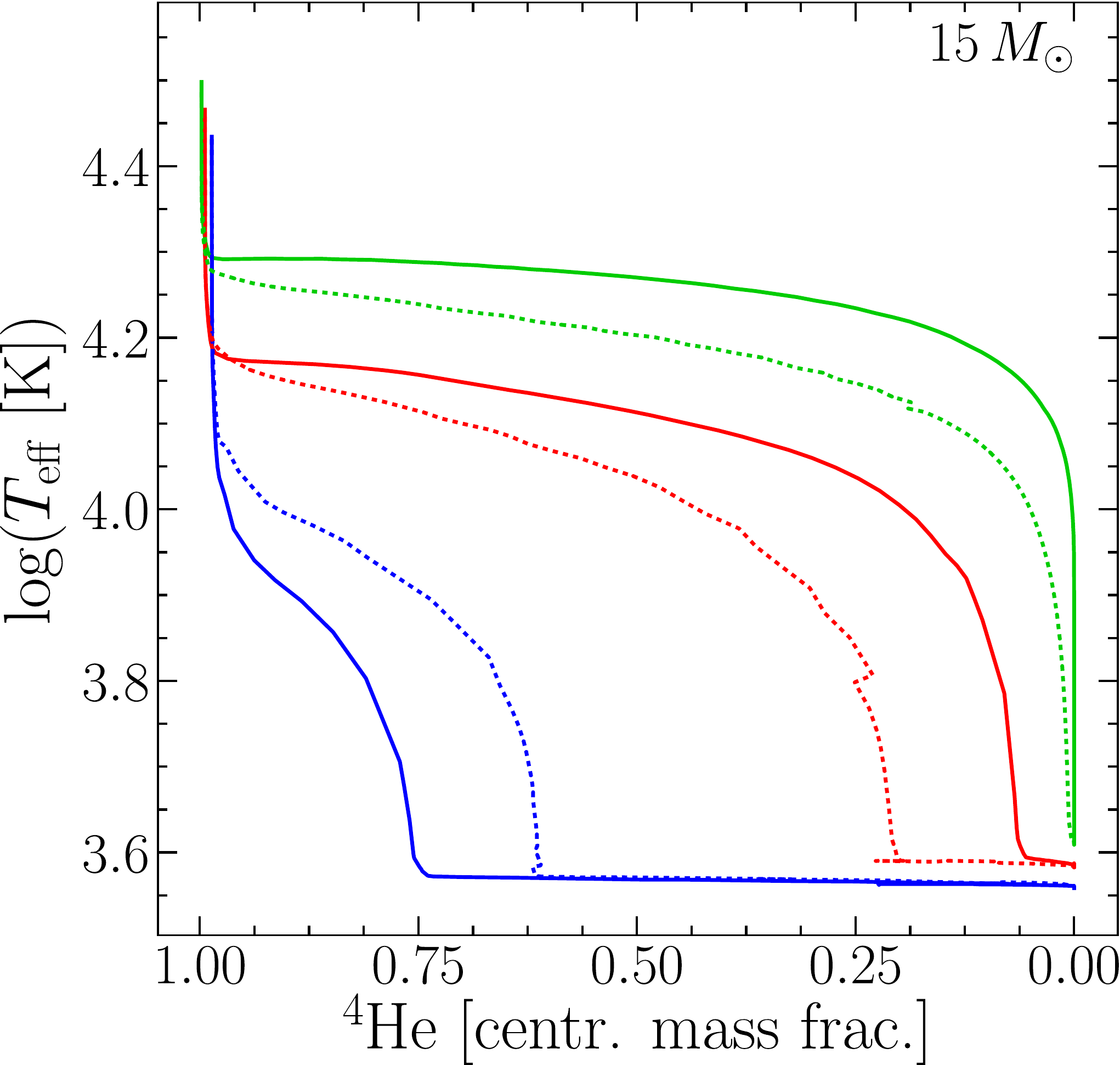}
\caption{Evolution of the effective temperature as a function of the mass fraction of helium at the center for 7, 9 and 15 M$_\odot$ models, with (solid lines) and without (dotted lines) rotation for different metallicities during the core He-burning phase. The blue, red and green lines indicate models with $Z=0.014$, 0.006 and 0.002, respectively (same colors as in Fig.~\ref{Fig:HRDcomp}).}
\label{Fig:teffy}
\end{figure*}

Figure~\ref{Fig:teffy} shows the evolution of the effective temperature during the core He-burning phase. These plots illustrate how fast a model crosses the HR gap after the MS and whether or not a blue loop occurs. The tracks corresponding to the Z=0.006 models (red lines) are in general framed by those corresponding to the Z=0.014 (blue lines) and Z=0.002 (green lines) models. 

A common feature to all these tracks is that they end their evolution as red supergiants. Important differences are found in the time when the model first becomes a red supergiant, the duration of the red supergiant phase and the presence or absence of a blue loop. In general, when rotation is included and/or the metallicity decreases, the fraction of the core He-burning lifetime spent in the blue increases. 

In the case of the 7 and 9 M$_\odot$ models, the star evolves rapidly to a red supergiant (RSG) phase after the MS and then a blue loop may or may not develop.
Rotation leads to longer blue loops in this mass range. Rotation makes larger helium cores, but also smooths the abundance gradients at the outer border of the H-burning shell.
Larger helium cores tend to decrease the extension of the blue loops \citep[][]{Lauterborn1971,MMVII2001}. On the other hand, the smoother variations of the abundances above the H-burning shell induced by rotation favor a blue loop. Two counteracting effects of rotation are thus at work here and we see that, in general, the smoothing effect of rotation dominates. Figure~\ref{Fig:teffy} also shows that the blue loops for the 7 and 9 M$_\odot$ models are more developed when the metallicity decreases. 

For the 15 M$_\odot$ models, no blue loops are present, but the beginning of the RSG phase is more and more delayed when the metallicity decreases. Let us recall that the crossing of the HR diagram is triggered by the ignition of the H-burning shell after the core contracts at the end of the H-burning phase. Some of the energy extracted from the gravitational reservoir in the core, in addition to energy released by the H-burning shell, is used to expand the envelope. Metallicity plays a role in modifying the hydrostatic structure of the models during and after the crossing as it affects the CNO content in the H-burning shell and the opacity of the outer layers. A lower metallicity favors core He ignition at a higher effective temperature. The timescale for crossing the HR gap from the Terminal Age Main Sequence to the red supergiant stage depends on many details of the models \citep[see for instance the discussion in][Farrell et al. in preparation]{Lauter1971, MMVII2001}. The differences between the behaviors of the 7, 9 and 15 M$_\odot$ shown in Fig.~\ref{Fig:teffy} indicates that lowering the metallicity has a different impact on the HR crossing for different initial masses, keeping everything else constant.

\subsection{Internal structures and lifetimes}

\begin{figure}
\includegraphics[width=0.238\textwidth]{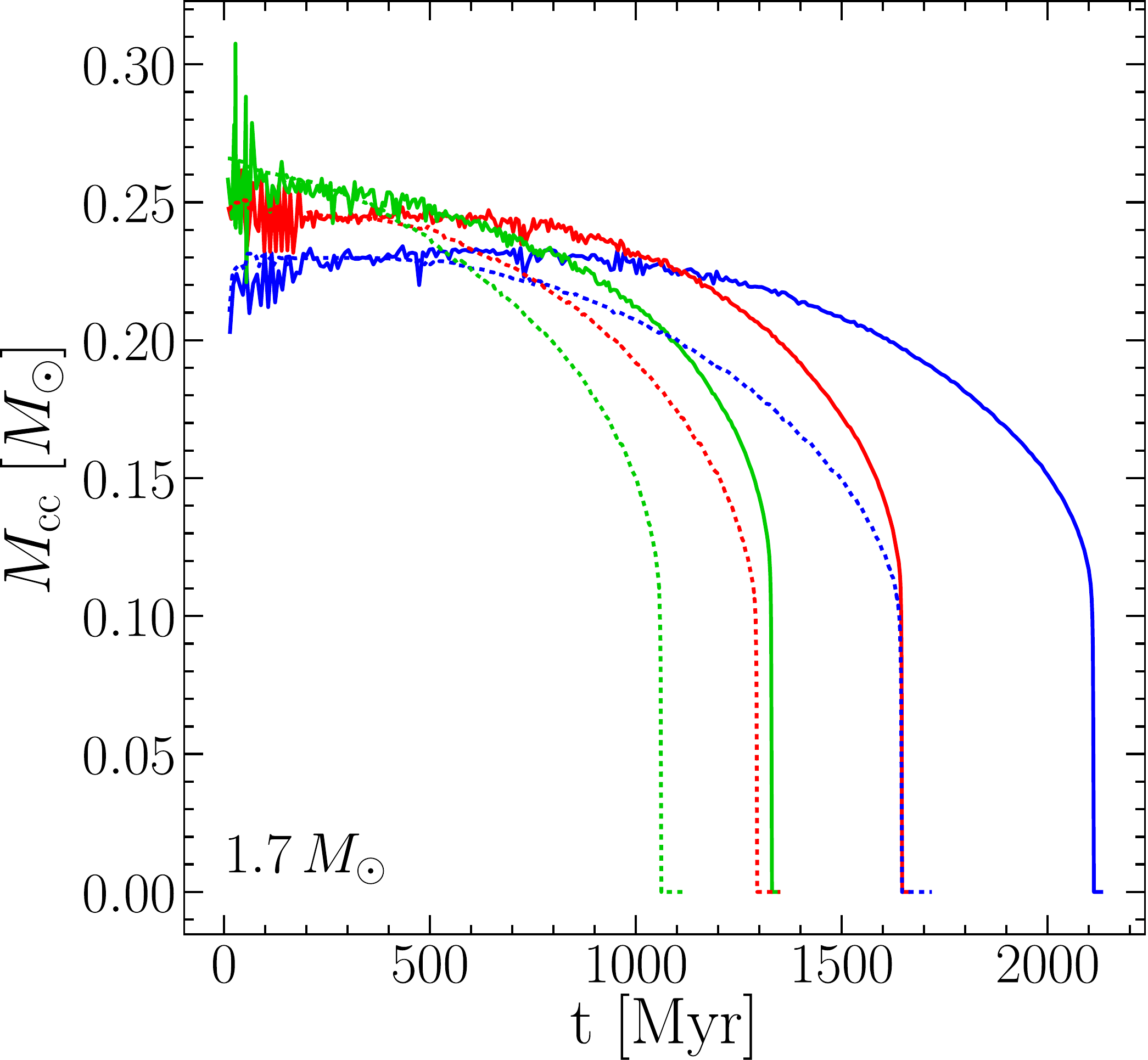}\includegraphics[width=0.23\textwidth]{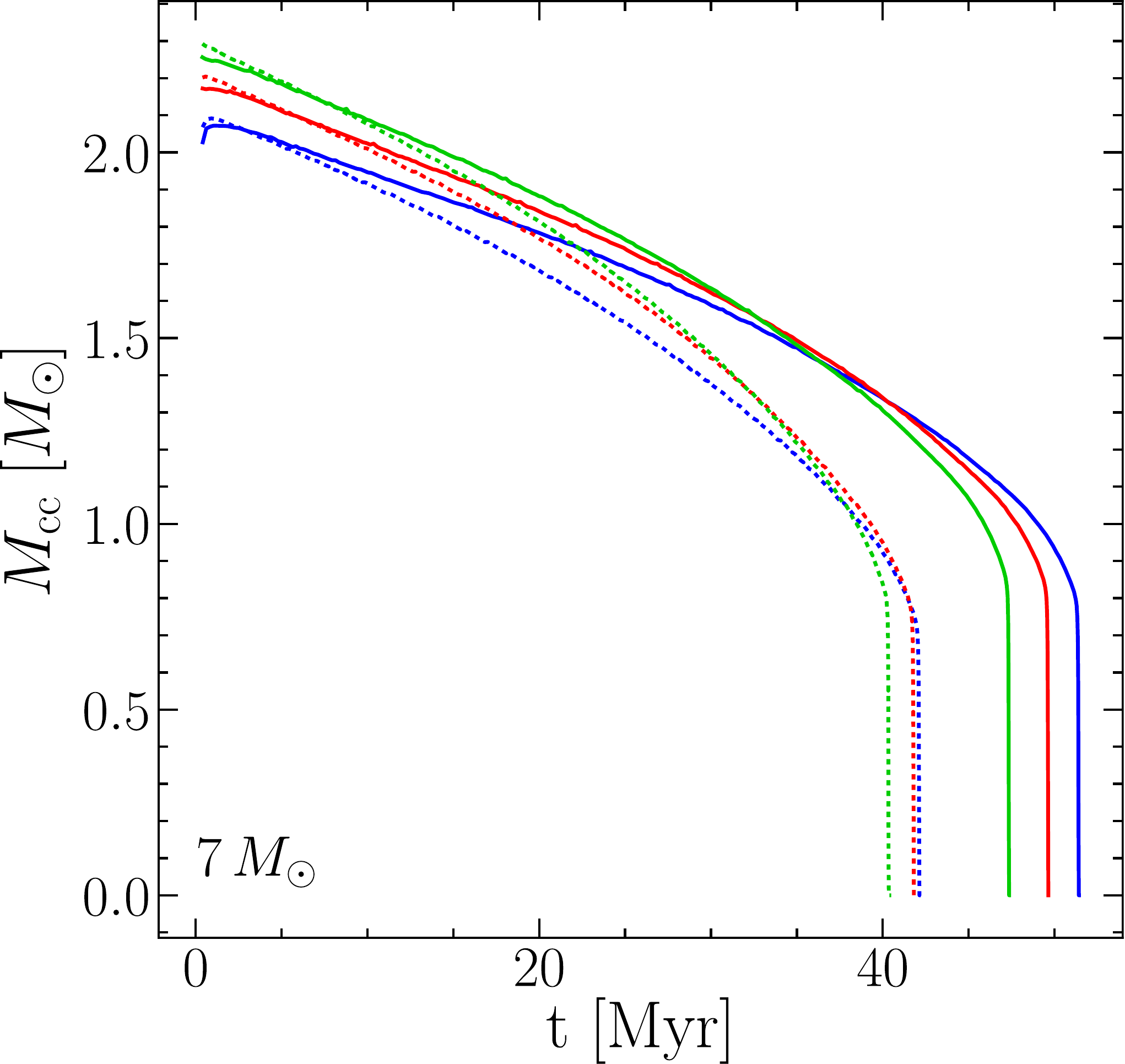}
\includegraphics[width=0.23\textwidth]{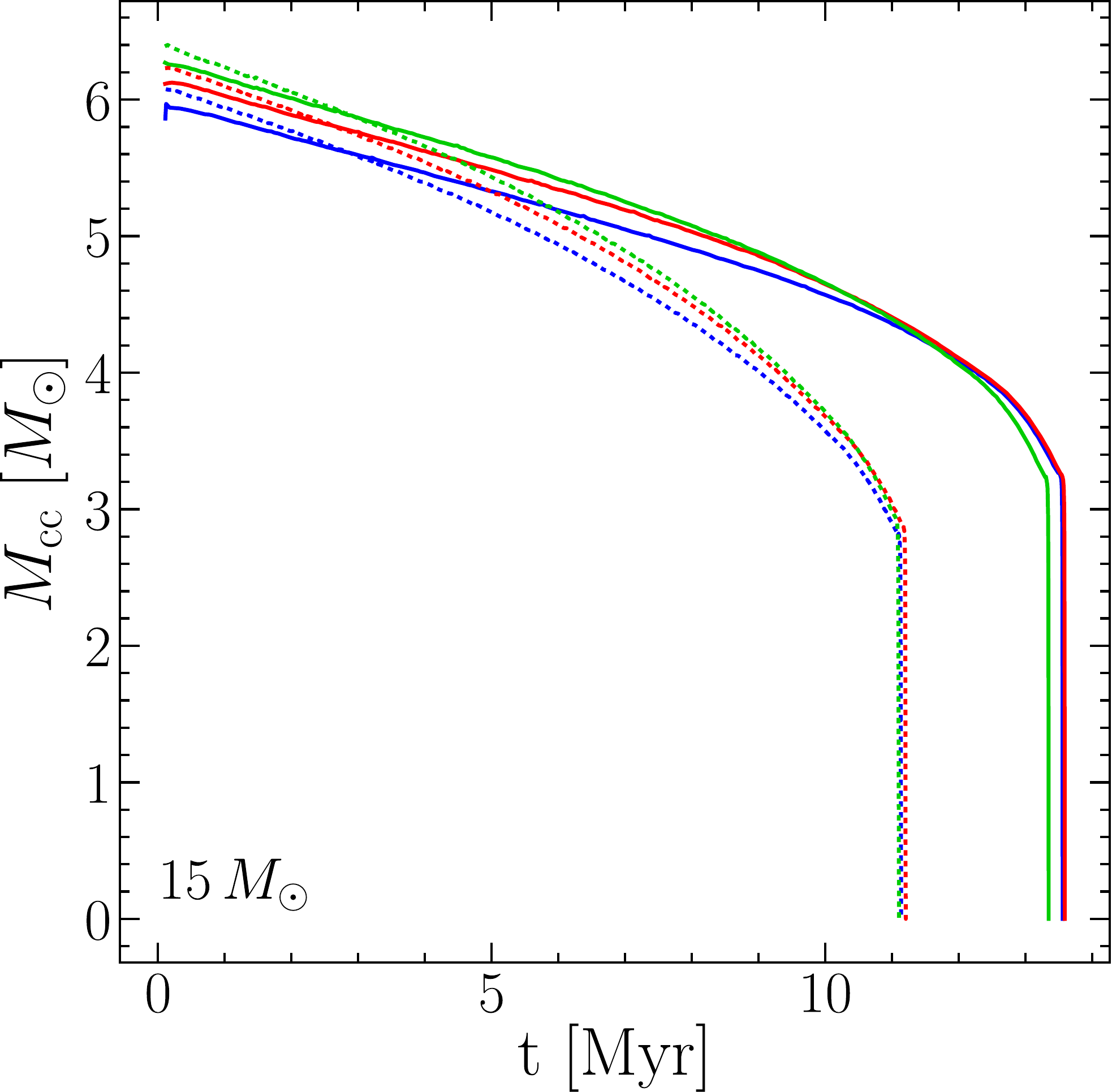}\includegraphics[width=0.238\textwidth]{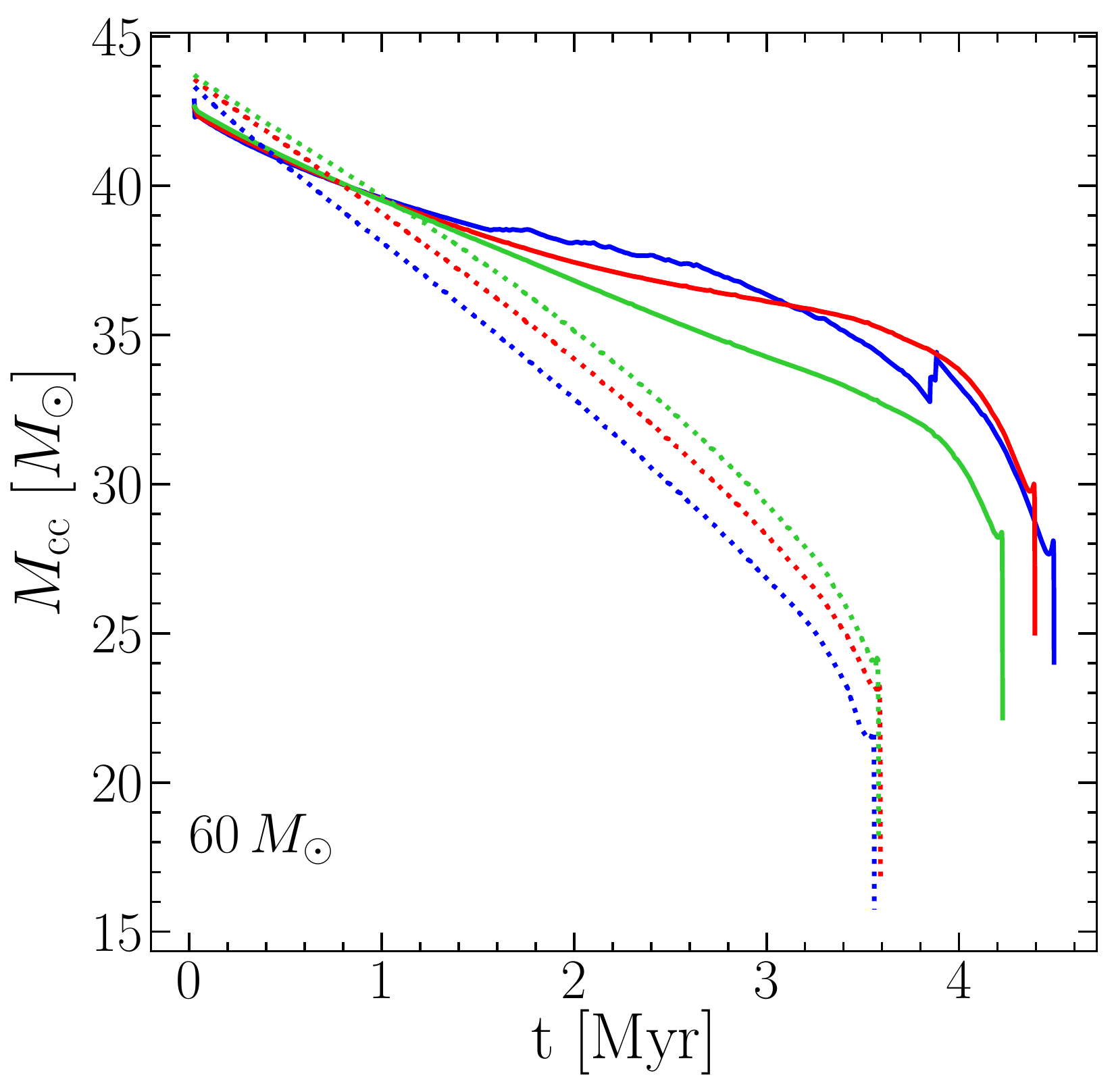}
\caption{Evolution of the convective core mass as a function of time during the MS phase for rotating (solid lines) and nonrotating (dotted lines) models, with 1.7, 7, 15 and 60 
M$_\odot$. The blue, red and green lines indicate models with $Z=0.014$, 0.006 and 0.002, respectively (same colors as in Fig.~\ref{Fig:HRDcomp}).
}
\label{Fig:convective}
\end{figure}

Figure~\ref{Fig:convective}  shows the evolution of the convective core during the H-burning phase for various stellar models as a function of time.
In the case of the nonrotating 7, 15 and 60 M$_\odot$ models, Fig.~\ref{Fig:convective} shows that a change in metallicity has no significant impact on the MS lifetime. 
This is no longer the case for the 1.7 M$_\odot$ models. The 1.7 M$_\odot$ model with Z=0.002 exhibits a MS lifetime that is shortened by 50\% with respect
to the MS lifetime of the corresponding Z=0.014 model. This is related to the fact that the luminosity is more affected by a change in metallicity for lower initial masses. For instance, changing the metallicity from 0.014 to 0.002 increases the luminosity by 60\% for a 1.7 M$_\odot$ model, while the corresponding increase in luminosity is only of about 20\% for the 7 M$_\odot$ model. A higher luminosity then reduces the MS lifetime.

Rotational mixing  increases the MS lifetime by continuously bringing fresh H-fuel in the core. An increase in the luminosity is however expected for rotating models because of the transport of helium in the envelope that globally decreases the opacity there. However, this occurs relatively late in the evolution, that is when the gradients of helium at the border of the core are sufficient\footnote{The diffusion velocity depends on the abundance gradient of the specific element considered \citep[see for instance Eq.~3 in][]{Meynet2004}.} and thus the addition of the fuel overcomes the effect linked to the luminosity increase. Rotation also slightly decreases the central temperature and density and thus
slows down the pace of the nuclear burning.

\begin{figure}
\includegraphics[width=0.24\textwidth]{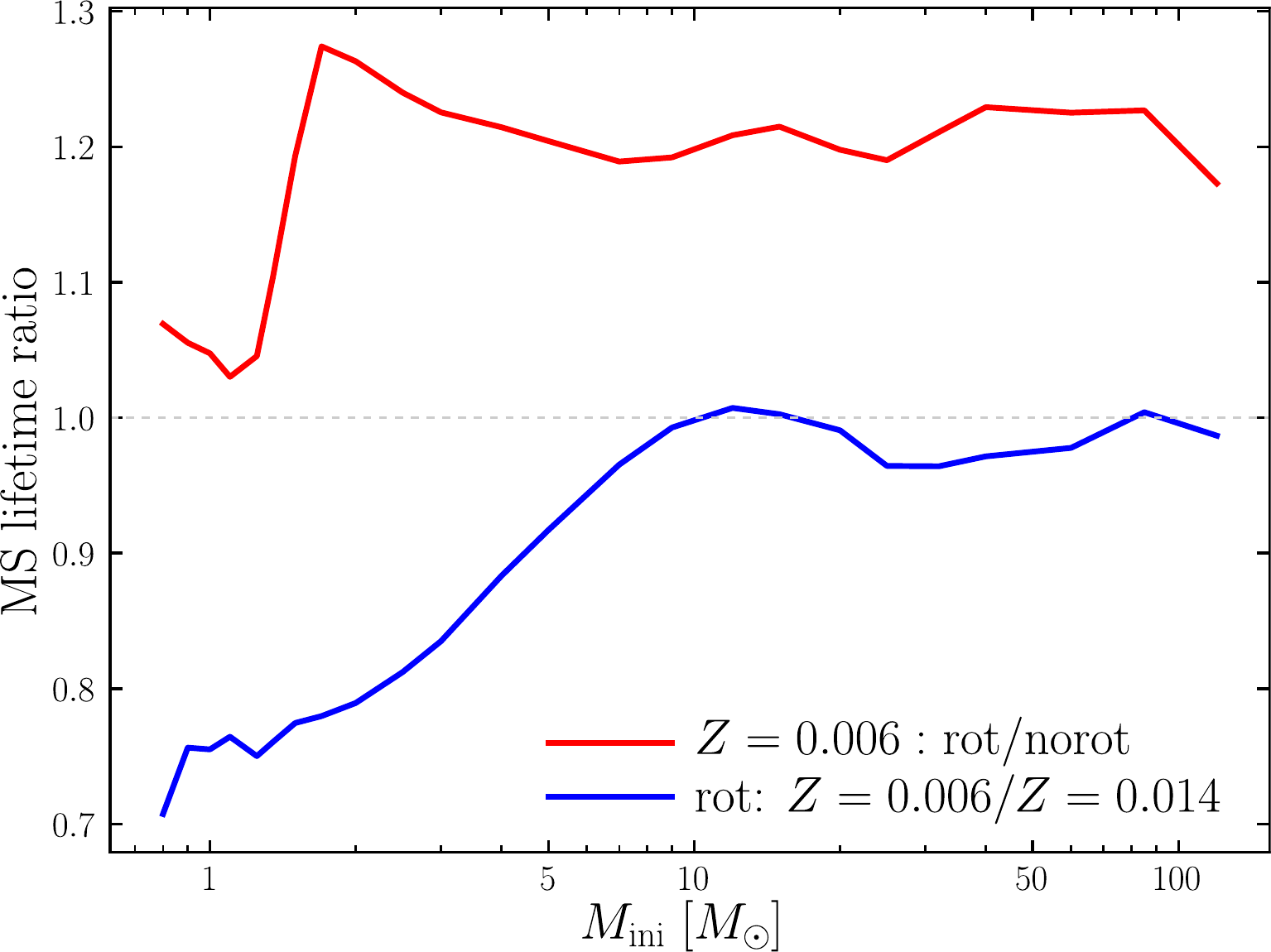}\includegraphics[width=0.24\textwidth]{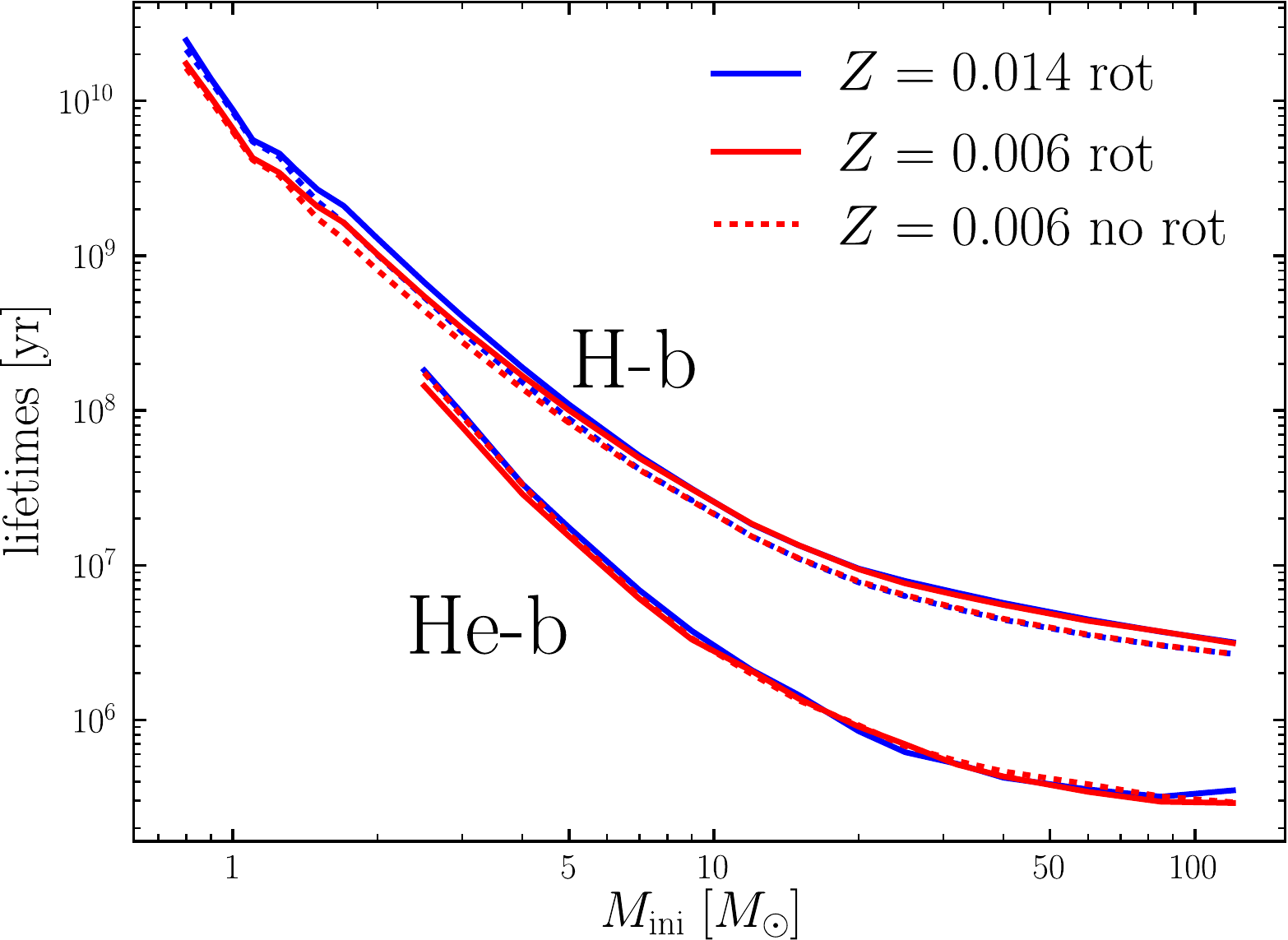}
\caption{
\textsl{Left panel:}  Ratio of the MS lifetimes between Z=0.006 rotating and non rotating models (red), and between rotating Z=0.006 and Z=0.014 models (blue) as a function of the initial mass.
\textsl{Right panel:} Lifetimes of the core H- and He-burning phases as a function of the initial mass for various metallicities.
}
\label{Fig:lifetimes}
\end{figure}

\begin{figure}
\includegraphics[width=0.24\textwidth]{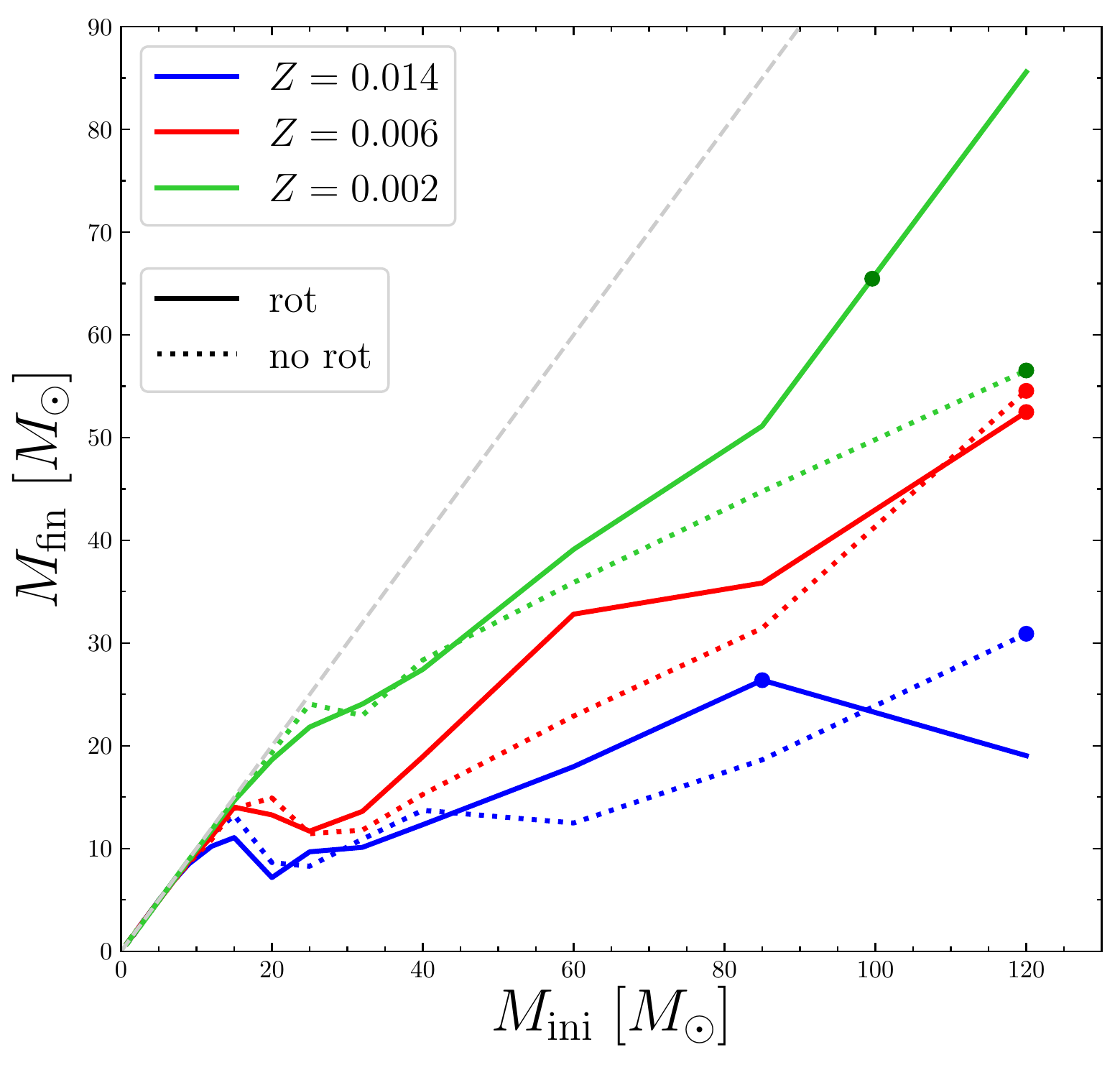}\includegraphics[width=0.24\textwidth]{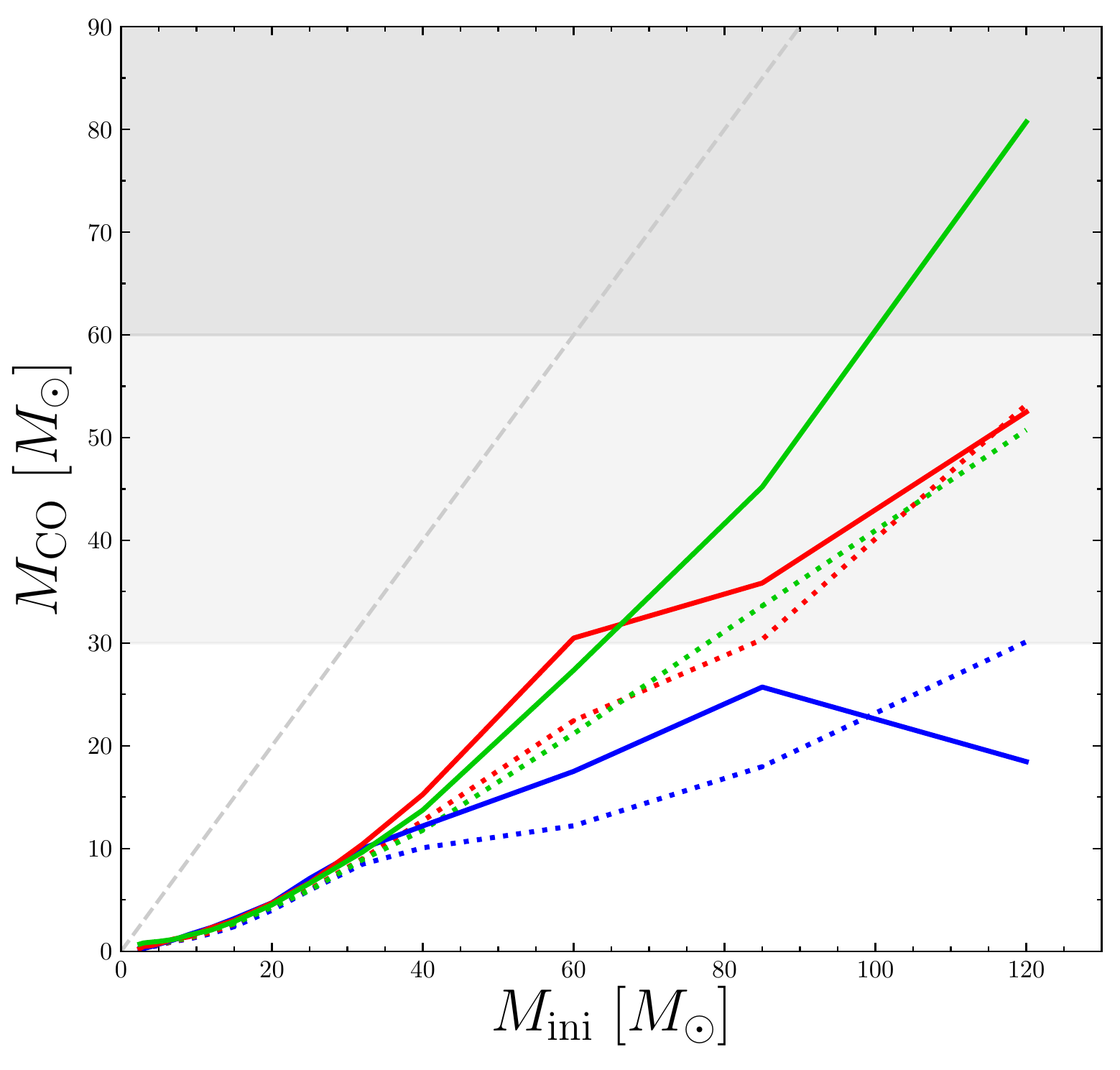}
\caption{\textsl{Left panel:} Variation of the final mass as a function of the initial mass for rotating (solid lines) and nonrotating (dotted lines) stellar models for various metallicities. The dot along each line indicates the probable maximum black hole mass that can be formed.
\textsl{Right panel:} Variation of the final Carbon-Oxygen core mass. The regions in light gray and in dark gray correspond to the domains where respectively pair pulsation instability and pair instability supernovae are expected to occur \citep[e.g.][]{Chatzopoulos2012, Woosley2017}.
}
\label{Fig:final}
\end{figure}

Figure~\ref{Fig:lifetimes} shows different lifetime ratios (left panel) and the lifetimes during the core H- and He-burning as a function of the initial mass for different metallicities. On average rotation increases the MS lifetime by about 20-25\% for masses between 1.7 and 120 M$_\odot$, and by 3-7\% for lower initial masses (see the red line in the left panel of Fig.~\ref{Fig:lifetimes}). In the low mass domain, magnetic braking removes a lot of angular momentum and thus considerably reduces the impact of rotational mixing over the whole MS phase. Comparing the MS lifetimes of the rotating Z=0.006 models to those for the metallicity Z=0.014 (blue line in the left panel of Fig.~\ref{Fig:lifetimes}), we see that shorter main sequence lifetimes are obtained for initial masses below about 8 M$_\odot$ for the lower metallicity models. This is mainly an effect of opacity. In less massive stars,
the opacity of the outer layers becomes less and less dominated by free electron scattering (as in the upper mass range), and more dominated by the opacity due to heavy elements. As a consequence, for these stars, when the metallicity decreases, the luminosity increases \citep[see for instance Fig. 1 in][]{ Nami1998} and this tends to shorten the MS lifetimes when compared to those of more metal rich stars. This effect is also obtained comparing nonrotating models \citep[see for instance Fig. 5 in][]{Nami1998}.

The right panel of Fig.~\ref{Fig:lifetimes} compares the absolute values of the lifetimes during the core H- and He-burning phases. Since the vertical scale is logarithmic, the differences seen on the left panel are barely visible. Globally, the ratio of the lifetimes between the core He and the H-burning lifetimes decreases when the mass increases. This trend is visible for the different metallicities as well as for models with and without rotation.

\subsection{Core masses}

Figure~\ref{Fig:final} shows the total final masses (left panel) and the final Carbon-Oxygen (CO, right panel) core masses as a function of the initial mass for models with and without rotation at different metallicities. As expected, the present $Z$=0.006 models fall inbetween the 0.002 and 0.014 models.

As is well known, a lower metallicity implies weaker mass losses by stellar winds that leads to larger final masses. 
In most cases, rotation tends to increase the final masses.
As seen before, rotating models spend a larger fraction of their lifetime in the blue part of the HRD where the mass losses are weaker.  
For the most massive stars at Z=0.014 and Z=0.006, rotating models exhibit however lower final masses. This comes from the fact that rotation
makes these models enter the Wolf-Rayet regime at an earlier stage, so that rotating models spend a larger fraction of their lifetime in a phase characterized by strong stellar winds.

Concerning the variation of the CO core mass\footnote{The CO core mass is defined as the mass inside the layer where the mass fraction of carbon+oxygen is larger than 0.75 and increases inward.} with the initial mass, we note that for the Z=0.014 models (blue lines), the upper parts of the curves are the same as the final masses.
This simply reflects that, for these models, stellar winds are efficient enough to produce naked CO cores. Comparing the nonrotating models at $Z$=0.006 (red dotted line) and 0.002 (green dotted line), we see that very similar curves are obtained. More differences are found in the case of rotating models. For masses below 60 M$_\odot$, the CO core masses are slightly larger in $Z$=0.006 models  than in models with $Z$=0.002, while it is the contrary in the higher mass range where mass loss plays a significant role.

The question of the formation of the most massive black holes has received some attention recently in particular following the gravitational wave detection GW190521 \citep[][]{abb20,Groh2020, Eoin2020}. On the left panel of Fig.~\ref{Fig:final}, a dot along each line indicates the maximum mass a black hole may have. Actually the values shown do not account for mass losses that can be induced by pulsational pair instabilities before the end of the evolution and thus represents some upper values. The point on the solid green line
corresponds to the final mass of the model that, on the right panel of  Fig.~\ref{Fig:final}, is just at the border between the light and dark gray regions. Models in the dark gray region
are expected to be completely destroyed by a pair instability supernova and thus will leave no black hole.
We see that the maximum black hole mass that can be obtained from present models is around 30 M$_\odot$ at $Z$=0.014, 55  M$_\odot$ at $Z$=0.006 and 65 M$_\odot$ at $Z$=0.002. According to \citet[][]{abb20b}, the masses of 89 BHs detected through their gravitational wave emission, have been determined. The lowest mass is 5.0$^{+1.4}_{-1.9}$ M$_\odot$ and the largest is 95.3$^{+28.7}_{-18.9}$ M$_\odot$. About half of the detected BHs (44) have masses between 5 and 30 M$_\odot$, half (43) between 30 and 70 M$_\odot$. At the moment 2 have masses above 80 M$_\odot$ clearly in the so called BH mass gap \citep[e.g.][]{Heger2003} due to the complete destruction of stars undergoing a  pair-instability supernova (PISN). Despite the fact that the evolution of the stars having given birth to these BHs is affected by close binary interactions, it is worth noting that mass loss by stellar winds, as implemented in the present models, allows the production of BHs above 30 M$_\odot$ for metallicities lower than solar. Indeed, metallicities found in the LMC and SMC are sufficiently low for stars to retain up to 55-65 M$_\odot$ until the late stages of their evolution. The exact BH masses that these stars would produce require detailed simulations of the PISN phase \citep[e.g.][]{2019ApJ...887...53F,2019ApJ...887...72L} but final BH masses are expected to be at least 40 M$_\odot$.

The right panel of Fig.~\ref{Fig:final} gives a first idea of the initial masses that at each metallicity will go through a phase of pulsational pair instability before to collapse to a black hole and those being completely destroyed by a pair instability \citep{Fowler1964, Rakavy1967, Barkat1967,Chatzopoulos2012, Woosley2017}. We see that the only models that would explode as a pair instability supernova would be the rotating models with an initial mass larger than 100 M$_\odot$ and a metallicity $Z$=0.002 in agreement with a previous work done using GENEC by \citet{Yusof2013}.

\section{Comparisons with observations}
\label{sec:obs}

The aim of this section is to illustrate the ability of the present models to reproduce some observational constraints.
This section will also give the opportunity to compare some outputs of the present stellar models with those obtained by other authors for a similar metallicity.

\subsection{Surface abundances on the Main-Sequence}

\begin{figure}
\begin{center}
\includegraphics[width=0.48\textwidth]{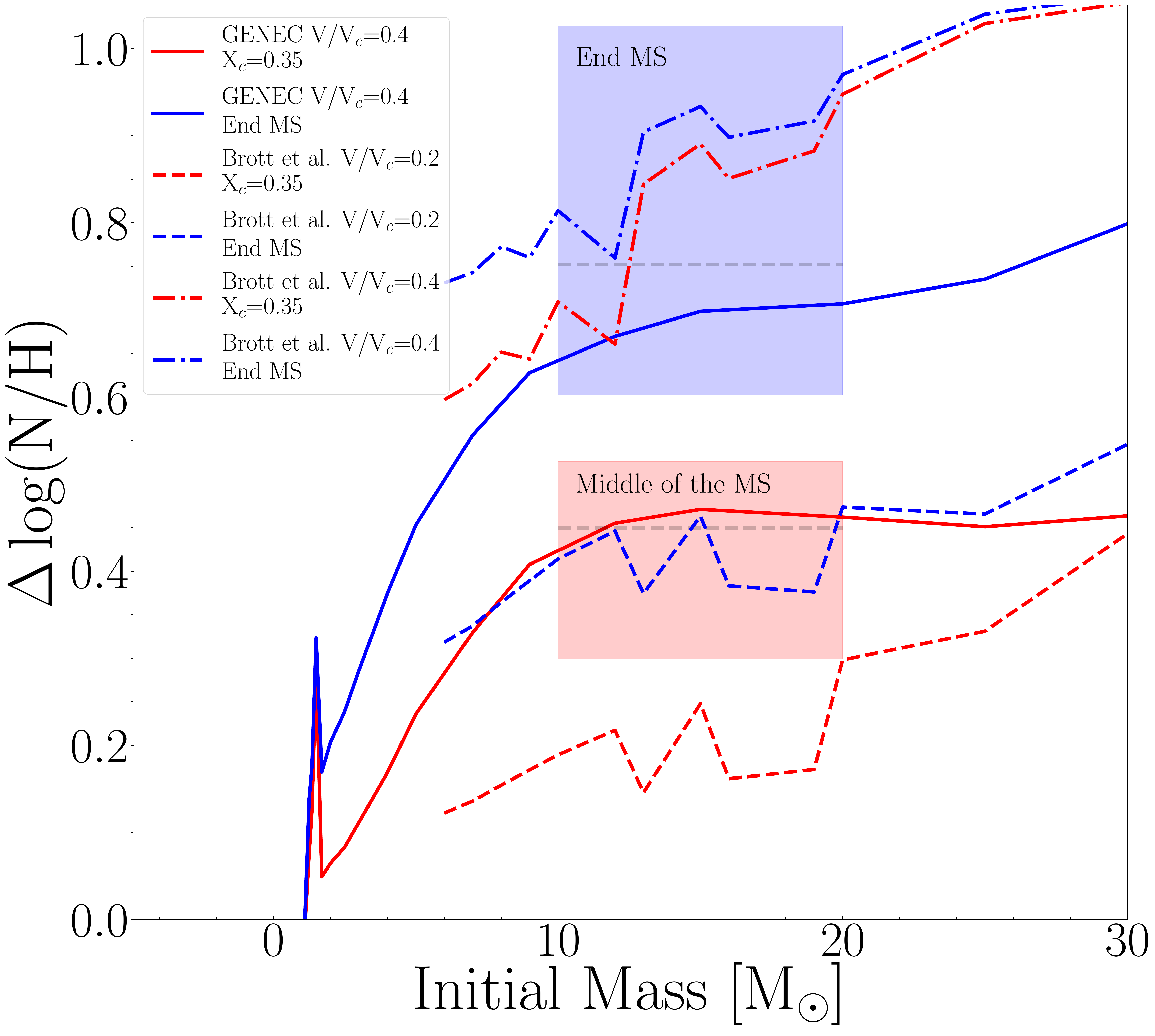}
\caption{
Values of the surface N/H ratio normalized to the initial one as a function of the initial mass of the models, in the middle of the MS ($X_{\rm c}=0.35$, red) and at the end of the MS ($X_{\rm c}=0.001$, blue). The shaded squares correspond to the location of the observed values from \citet{Dufton2018} in this diagram.
}
\label{Fig:abundMS}
\end{center}
\end{figure}

Figure~\ref{Fig:abundMS} shows the values of the N/H surface abundance ratio normalized to the same surface abundance ratio at the ZAMS, as a function of the initial mass, at two different stages during the MS. We recall here that rotational mixing is due to turbulence induced by various instabilities. These turbulent motions cannot be followed in one dimensional rotating models so that there is always one free parameter, which is typically related to the uncertain modeling of horizontal turbulence in the framework of the shellular rotation formalism developed by \citet{Zahn1992} that needs to be adjusted \citep[e.g.][]{Eggenberger2008}.
This parameter is calibrated in order for typical massive stars, at solar metallicity, with masses between 10 and 20 M$_\odot$, showing a time-averaged surface rotation velocity around 180-200 km s$^{-1}$ on the MS to reproduce the observed surface enrichments. This calibration, for the present models, is discussed in \citet{Ekstrom2012a}. This calibrated value has then been kept constant for all other metallicities \citep{Georgy2013a, groh2019} and also for the present Z=0.006 grid. Figure~\ref{Fig:abundMS}
allows to check the ability of the present rotating models, calibrated at solar metallicity, to reproduce the average surface enrichments observed at the metallicity of the LMC. This figure also compares the present results for the surface enrichments during the MS phase with those obtained by \citet{Brott2011}.

We first discuss the predictions of the present models. The trend obtained at the LMC metallicity is qualitatively similar to the one obtained at solar metallicity by \citet[][see their Fig.~11 and the interested reader is invited to read the subsection 5.4 in this paper for more details]{Ekstrom2012a}. The main difference is that, for a mass around 15 M$_\odot$, for a given value of the ratio between the surface rotation at the ZAMS and the critical velocity, the surface enrichment is larger at the LMC metallicity both in the middle of the MS phase and at the end. At solar metallicity, values for $\Delta \log (N/H)$ equal to $\sim$0.32 ($X_{\rm c}=0.35$) and $\sim$0.61 ($X_{\rm c}=0.001$) are found. The corresponding values at $Z$=0.006 are 0.46 and 0.70. We note however that the values for the 30 M$_\odot$ models at $Z$=0.006 are lower than at $Z$=0.014. This is due to the fact that, in this mass range, mass loss by stellar winds also play an important role at $Z$=0.014 in changing the surface abundances by unveiling inner stellar layers.

The red hatched zone in Fig.~\ref{Fig:abundMS} shows the region where the slightly enriched stars in the middle of the MS phase are observed. 
The blue one shows where the highly enriched stars at the end of the MS phase are observed. The observations are taken from \citet{Dufton2018}.
We see that the present models fit the averaged surface abundance ratio in the middle of the MS phase quite well. 
They also provide a good fit to the observed values at the end of the MS phase, however slightly below the averaged value.  
The models by \citet{Brott2011} with initial rotation rates $\upsilon/\upsilon_{\rm crit}=0.2$ and 0.4 show slightly lower and higher surface enrichments than the present models, respectively (see dashed and dotted-dashed lines in Fig.~\ref{Fig:abundMS}). This likely is due to differences in the way of implementing rotational mixing in the two grids of models. Despite the fact that both grids are calibrated in a very similar way at solar metallicity, the different implementations of chemicals and angular momentum transport do not provide identical results at other metallicities. A detailed discussion about the reasons of these differences is beyond the scope of the present paper. Let us just mention here that the models by \citet{Brott2011} take into account magnetic effects for the internal transport of angular momentum and that the implementation of the transport by meridional currents as well as the way of accounting for the inhibiting effects of chemical composition gradients differ between the two grids of models.

\begin{figure}
\includegraphics[width=0.48\textwidth]{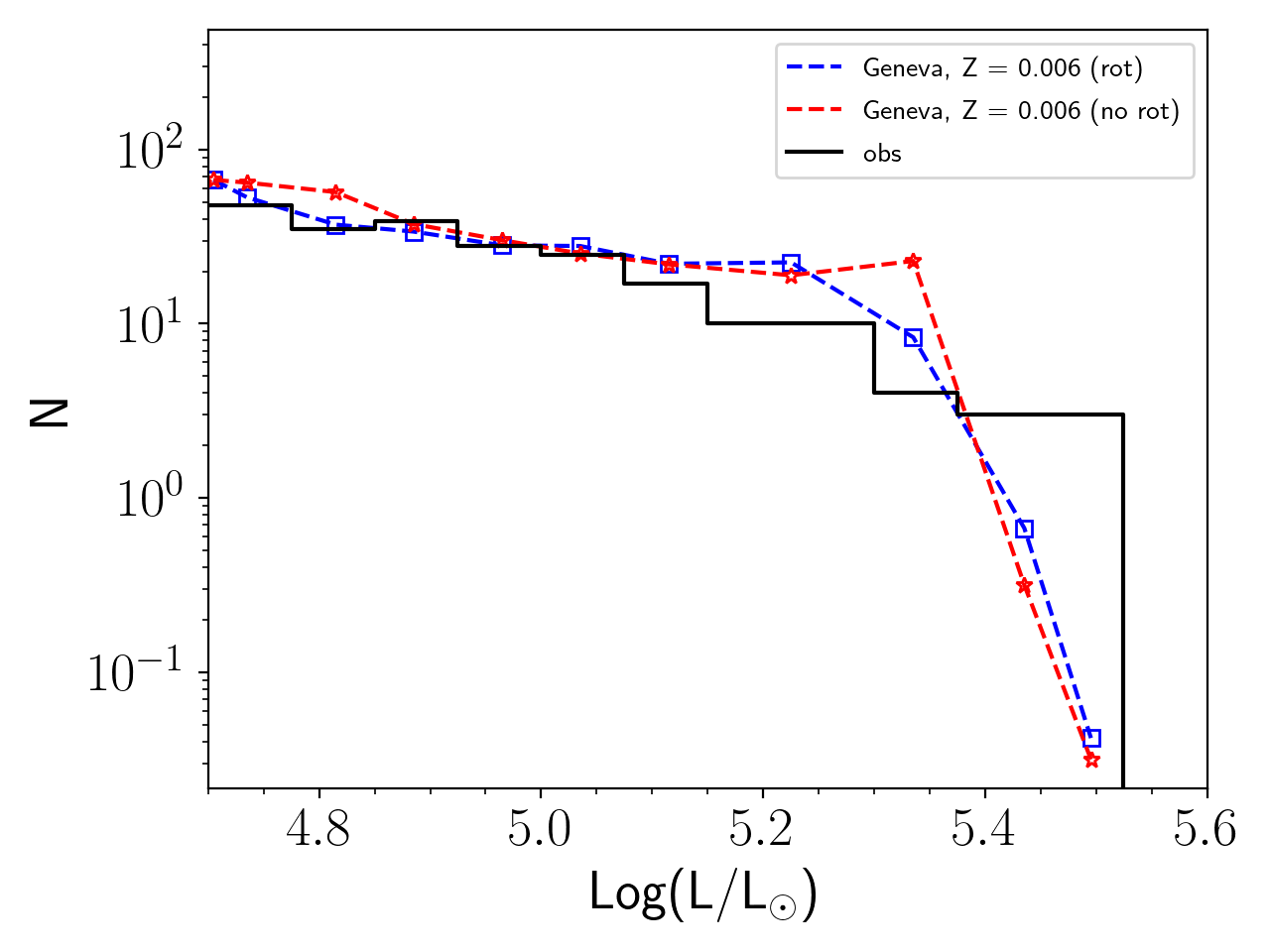}
\caption{Luminosity function of red supergiants in the LMC. The
black histogram shows the observed luminosity function as given in \citet[][and see references therein]{Davies2018} and using observations by \citet{Elias1985, vanLoon2005, Buchanan2006, Bonanos2009, Bonanos2010, Neugent2010, Neugent2012, Gonzalez2015, Goldman2017}.  
The results of the population synthesis code SYCLIST  \citep{Georgy2014} using the present models at Z=0.006 are shown for rotating (blue squares) and non rotating (red stars) models.
The numbers of stars in the models have been normalized to reproduce the observed number of stars at a luminosity equal to $\log L/{\rm L}_\odot$=5.0.
}
\label{Fig:lum}
\end{figure}

\subsection{Red supergiants}

The slope of the observed luminosity function of red supergiants is an interesting feature to compare the predictions of stellar models to. This slope is affected by the time-averaged mass loss rate during the red supergiant phase, becoming steeper for stronger mass losses \citep{Meynet2015}. This time-averaged mass loss rate encompasses both the quiescent low mass loss rate phase of red supergiants and
also the short outbursts during which important amounts of mass can be lost and that are more difficult to catch observationally. Recently, \citet{Neugent2020} using the models by \citet{Ekstrom2012a} at $Z$=0.014
concluded that these models provide an excellent fit to the luminosity function of red supergiants observed in M31.  In Fig.~\ref{Fig:lum}, we compare the red supergiant luminosity function obtained by the present
models at $Z$=0.006 with the observed one given by  \citet[][and see the detailed references for the observations in the figure caption]{Davies2018}. The model predictions have been obtained assuming a constant star formation rate and a Salpeter's initial mass function. The predictions have been normalized to reproduce the observed luminosity function obtained at a luminosity $\log L/{\rm L_\odot}=5.0$. We chose the bin around this luminosity for normalizing the population synthesis models, because this bin contains a significant number of stars and is in the middle of the slowly varying part of the luminosity function. These two features give some confidence about the robustness of the observed trend in that part of the luminosity function. Choosing another bin in that same region for the normalization would not significantly change the results. The agreement between the theory and observation is very good for the luminosity range between 4.7 and 5.15. Above that luminosity, but below 5.4, models predict significantly more red supergiants than observed (especially the nonrotating models), while beyond 5.4, the reverse situation occurs. In the last two observed bins (for $\log L/{\rm L_\odot} > 3$), however, the number of stars is very low (3-4 stars), thus stochastic effects can blur the picture. There is also one red supergiant at $\log L/{\rm L_\odot}=5.8$ (outside the frame of the figure) that would not be matched by the present model, but again stochastic effects may be responsible for that too.

On the whole, the comparison shows that the models do a reasonable job in fitting the observed red supergiant luminosity function at least in the part where it is well populated. The agreement is less good, as noted above, in the part where the luminosity function shows a fast decrease. It is however not possible at that stage to know whether this disagreement is linked to a weakness of the models or to some stochastic effects.

\section{Conclusions}
\label{sec:conclusion}

The present grid of stellar models complements the grids at $Z=0.014$, 0.002, 0.0004 and  $Z=0$  \citep[][]{Ekstrom2012a, Georgy2012, Georgy2013a, groh2019, Murphy2020}. Together these models allow the exploration of the impact of changing the metallicity, the initial mass and the initial composition over large domains, keeping the other ingredients of the stellar models constant. 
The main conclusions of the present work are indicated below:
\begin{itemize}
\item The properties of the $Z$=0.006 models fall in general inbetween the properties of the models with $Z$ equal to 0.002 and 0.014.
\item The present rotating models can reproduce the averaged surface nitrogen enrichments observed at the surface of B-type stars in the Large Magellanic Cloud.
\item They fit reasonably well the slope of the luminosity function of the red supergiants in the LMC. 
\item The present single star models indicate that the most massive black holes that can be obtained at $Z$ equal to 0.014, 0.006 and 0.002 are around 30, 55 and 65 M$_\odot$, respectively.
Only rotating models at $Z$=0.002 and with initial masses larger than $\sim$ 100 M$_\odot$ form massive enough carbon-oxygen cores to enter into the pair instability supernova regime.
\end{itemize}
The numerical data of the present grid are directly accessible on the web page:
{\scriptsize https://www.unige.ch/sciences/astro/evolution/fr/base-de-donnees/}.

\begin{acknowledgements} 
We thank the referee for useful comments. PE, SE, CG, DN, GM, SS and LH have received funding from the European Research Council (ERC) under the European Union's Horizon 2020 research and innovation programme (grant agreement No 833925, project STAREX). The work done by SM and CP is supported by the Swiss National Science Foundation grant 200020-172505. GB acknowledges fundings from the SNF AMBIZIONE grant No 185805 (Seismic inversions and modeling of transport processes in stars). RH acknowledges support from the “ChETEC” COST Action (CA16117) and the IReNA AccelNet Network of Networks (NSF Grant No. OISE-1927130).
\end{acknowledgements}

\appendix
\section{Main properties of the stellar models}

Table~\ref{TabListModels} presents the main properties of the stellar models on the ZAMS, at the end of the core H-, He- and C-burning (when it applies). The first three columns give the initial mass, the initial rotation velocity at the equator and the time-averaged rotation velocity during the MS. Columns 4 to 9 present respectively the duration of the core H-burning phase, the actual mass, the surface rotation velocity, the surface mass fraction of helium, and the surface nitrogen to carbon, and nitrogen to oxygen mass ratios at the end of the core H-burning phase. Columns 10 to 15 and 16 to 21 indicate the same physical quantities as those indicated in columns 4 to 9, but at the end of the core He-burning phase and at the end of the core C-burning phase, respectively. 
\begin{landscape}
\begin{table}
\caption{Properties of the $Z=0.006$ models at the end of the H-, He-, and C-burning phases.}
\centering
\scalebox{0.7}{
\begin{tabular}{rrr|rrrrrr|rrrrrr|rrrrrr}
\hline\hline
\multicolumn{3}{c|}{} & \multicolumn{6}{c|}{End of H-burning} & \multicolumn{6}{c|}{End of He-burning} & \multicolumn{6}{c}{End of C-burning}\\
$M_\text{ini}$ & $v_\text{eq}$ & $\bar{v}_\text{MS}$ & $\tau_\text{H}$ & $M$ & $v_\text{eq}$ & $Y_\text{surf}$ & $\text{N}/\text{C}$ & $\text{N}/\text{O}$ & $\tau_\text{He}$ & $M$ & $v_\text{eq}$ & $Y_\text{surf}$ & $\text{N}/\text{C}$ & $\text{N}/\text{O}$ & $\tau_\text{C}$ & $M$ & $v_\text{eq}$ & $Y_\text{surf}$ & $\text{N}/\text{C}$ & $\text{N}/\text{O}$ \\
$M_{\sun}$ & \multicolumn{2}{c|}{km s$^{-1}$} & Myr & $M_{\sun}$ & km s$^{-1}$ & \multicolumn{3}{c|}{mass fract.} & Myr & $M_{\sun}$ & km s$^{-1}$ & \multicolumn{3}{c|}{mass fract.} & kyr & $M_{\sun}$ & km s$^{-1}$ & \multicolumn{3}{c}{mass fract.}\\
\hline
$120.00$ & $  0.$ & $  0.$ & $    2.674$ & $ 79.08$ & $  0.$ & $0.5893$ & $106.2743$ & $ 78.1921$ & $    0.294$ & $ 54.63$ & $  0.$ & $0.2237$ & $  0.0000$ & $  0.0000$ & $    0.002$ & $ 54.55$ & $  0.$ & $0.2236$ & $  0.0000$ & $  0.0000$ \\
$120.00$ & $435.$ & $204.$ & $    3.136$ & $ 64.34$ & $  9.$ & $0.9873$ & $ 60.5208$ & $108.8550$ & $    0.290$ & $ 41.88$ & $  0.$ & $0.2988$ & $  0.0000$ & $  0.0000$ & $    0.004$ & $ 41.80$ & $  0.$ & $0.2960$ & $  0.0000$ & $  0.0000$ \\
$ 85.00$ & $  0.$ & $  0.$ & $    3.040$ & $ 57.76$ & $  0.$ & $0.4716$ & $126.5064$ & $ 72.6224$ & $    0.320$ & $ 31.53$ & $  0.$ & $0.2819$ & $  0.0000$ & $  0.0000$ & $    0.007$ & $ 31.45$ & $  0.$ & $0.2746$ & $  0.0000$ & $  0.0000$ \\
$ 85.00$ & $410.$ & $207.$ & $    3.730$ & $ 59.52$ & $ 16.$ & $0.9796$ & $ 64.0934$ & $104.4296$ & $    0.296$ & $ 37.71$ & $  0.$ & $0.2887$ & $  0.0000$ & $  0.0000$ & $    0.005$ & $ 37.62$ & $  0.$ & $0.2846$ & $  0.0000$ & $  0.0000$ \\
$ 60.00$ & $  0.$ & $  0.$ & $    3.563$ & $ 54.96$ & $  0.$ & $0.2559$ & $  0.2885$ & $  0.1152$ & $    0.384$ & $ 22.92$ & $  0.$ & $0.1469$ & $  0.0004$ & $  0.0003$ & $    0.008$ & $ 22.90$ & $  0.$ & $0.1465$ & $  0.0004$ & $  0.0003$ \\
$ 60.00$ & $378.$ & $217.$ & $    4.366$ & $ 49.24$ & $ 14.$ & $0.6102$ & $ 17.0903$ & $  6.7100$ & $    0.343$ & $ 33.76$ & $  0.$ & $0.8168$ & $143.6148$ & $ 50.2052$ & $    0.004$ & $ 33.42$ & $  0.$ & $0.8436$ & $127.7616$ & $ 66.8659$ \\
$ 40.00$ & $  0.$ & $  0.$ & $    4.503$ & $ 37.97$ & $  0.$ & $0.2559$ & $  0.2885$ & $  0.1152$ & $    0.465$ & $ 15.55$ & $  0.$ & $0.6555$ & $145.5090$ & $ 69.1625$ & $    0.047$ & $ 15.27$ & $  0.$ & $0.9941$ & $ 46.2786$ & $129.6218$ \\
$ 40.00$ & $334.$ & $234.$ & $    5.535$ & $ 36.31$ & $ 31.$ & $0.4307$ & $  5.6063$ & $  1.8030$ & $    0.431$ & $ 19.75$ & $  0.$ & $0.7214$ & $157.2104$ & $ 25.4387$ & $    0.020$ & $ 19.39$ & $  0.$ & $0.7324$ & $144.2723$ & $ 35.6081$ \\
$ 32.00$ & $  0.$ & $  0.$ & $    5.286$ & $ 30.90$ & $  0.$ & $0.2559$ & $  0.2885$ & $  0.1152$ & $    0.543$ & $ 12.53$ & $  0.$ & $0.5848$ & $170.4232$ & $ 54.1874$ & $    0.123$ & $ 11.81$ & $  0.$ & $0.7445$ & $108.1579$ & $ 74.4552$ \\
$ 32.00$ & $334.$ & $242.$ & $    6.401$ & $ 30.21$ & $103.$ & $0.3486$ & $  3.7147$ & $  1.0528$ & $    0.515$ & $ 14.54$ & $  2.$ & $0.6776$ & $169.1095$ & $ 18.1632$ & $    0.070$ & $ 14.00$ & $  0.$ & $0.6815$ & $137.3723$ & $ 20.8927$ \\
$ 25.00$ & $  0.$ & $  0.$ & $    6.403$ & $ 24.52$ & $  0.$ & $0.2559$ & $  0.2885$ & $  0.1152$ & $    0.677$ & $ 12.05$ & $  0.$ & $0.5601$ & $ 93.2948$ & $ 27.4789$ & $    0.364$ & $ 11.45$ & $  0.$ & $0.5669$ & $114.7408$ & $ 39.0823$ \\
$ 25.00$ & $301.$ & $242.$ & $    7.621$ & $ 24.28$ & $182.$ & $0.3069$ & $  3.2674$ & $  0.7865$ & $    0.696$ & $ 12.57$ & $  0.$ & $0.5830$ & $ 65.5787$ & $  9.6140$ & $    0.223$ & $ 11.84$ & $  0.$ & $0.5899$ & $ 93.6585$ & $ 12.1617$ \\
$ 20.00$ & $  0.$ & $  0.$ & $    7.862$ & $ 19.80$ & $  0.$ & $0.2559$ & $  0.2885$ & $  0.1152$ & $    0.924$ & $ 15.33$ & $  0.$ & $0.3312$ & $  3.3949$ & $  0.7816$ & $    0.755$ & $ 14.93$ & $  0.$ & $0.4012$ & $  7.1974$ & $  1.5900$ \\
$ 20.00$ & $292.$ & $234.$ & $    9.418$ & $ 19.71$ & $227.$ & $0.2911$ & $  3.4210$ & $  0.7065$ & $    0.904$ & $ 12.90$ & $  1.$ & $0.5360$ & $ 42.3704$ & $  5.0983$ & $    0.829$ & $ 12.36$ & $  0.$ & $0.5528$ & $ 61.7054$ & $  6.9411$ \\
$ 15.00$ & $  0.$ & $  0.$ & $   11.097$ & $ 14.87$ & $  0.$ & $0.2559$ & $  0.2885$ & $  0.1152$ & $    1.338$ & $ 14.00$ & $  0.$ & $0.2685$ & $  1.3634$ & $  0.3566$ & $    3.504$ & $ 13.90$ & $  0.$ & $0.3214$ & $  2.7453$ & $  0.6739$ \\
$ 15.00$ & $271.$ & $213.$ & $   13.482$ & $ 14.80$ & $196.$ & $0.2835$ & $  3.6913$ & $  0.6771$ & $    1.374$ & $ 13.97$ & $  1.$ & $0.3590$ & $  8.4824$ & $  1.1263$ & $    1.885$ & $ 13.85$ & $  1.$ & $0.3883$ & $ 10.8169$ & $  1.3220$ \\
$ 12.00$ & $  0.$ & $  0.$ & $   15.310$ & $ 11.95$ & $  0.$ & $0.2559$ & $  0.2885$ & $  0.1152$ & $    1.976$ & $ 10.95$ & $  0.$ & $0.2667$ & $  1.5412$ & $  0.3811$ & $    6.465$ & $ 10.88$ & $  0.$ & $0.2977$ & $  2.4608$ & $  0.5905$ \\
$ 12.00$ & $258.$ & $205.$ & $   18.502$ & $ 11.93$ & $206.$ & $0.2751$ & $  3.3683$ & $  0.6202$ & $    2.066$ & $ 11.26$ & $  1.$ & $0.3175$ & $  7.1470$ & $  0.9174$ & $    3.613$ & $ 11.17$ & $  1.$ & $0.3560$ & $  9.5810$ & $  1.1221$ \\
\cline{16-21}
$  9.00$ & $  0.$ & $  0.$ & $   25.992$ & $  8.99$ & $  0.$ & $0.2559$ & $  0.2885$ & $  0.1152$ & $    3.363$ & $  8.65$ & $  0.$ & $0.2614$ & $  1.5134$ & $  0.3785$ &  \\
$  9.00$ & $264.$ & $195.$ & $   30.985$ & $  8.99$ & $204.$ & $0.2692$ & $  2.7550$ & $  0.5504$ & $    3.307$ & $  8.80$ & $  1.$ & $0.3101$ & $  7.7002$ & $  0.8966$ &  \\
$  7.00$ & $  0.$ & $  0.$ & $   41.395$ & $  7.00$ & $  0.$ & $0.2559$ & $  0.2885$ & $  0.1152$ & $    6.275$ & $  6.89$ & $  0.$ & $0.2587$ & $  1.3783$ & $  0.3438$ &  \\
$  7.00$ & $247.$ & $185.$ & $   49.222$ & $  7.00$ & $194.$ & $0.2644$ & $  1.9348$ & $  0.4539$ & $    6.074$ & $  6.90$ & $  2.$ & $0.3044$ & $  6.4889$ & $  0.8419$ &  \\
$  5.00$ & $  0.$ & $  0.$ & $   83.169$ & $  5.00$ & $  0.$ & $0.2559$ & $  0.2885$ & $  0.1152$ & $   15.908$ & $  4.95$ & $  0.$ & $0.2582$ & $  1.3024$ & $  0.3256$ &  \\
$  5.00$ & $215.$ & $173.$ & $  100.156$ & $  5.00$ & $178.$ & $0.2607$ & $  1.2423$ & $  0.3473$ & $   15.358$ & $  4.93$ & $  3.$ & $0.2977$ & $  4.8488$ & $  0.7615$ &  \\
$  4.00$ & $  0.$ & $  0.$ & $  137.775$ & $  4.00$ & $  0.$ & $0.2559$ & $  0.2885$ & $  0.1152$ & $   33.275$ & $  3.96$ & $  0.$ & $0.2623$ & $  1.4460$ & $  0.3596$ &  \\
$  4.00$ & $219.$ & $165.$ & $  167.317$ & $  4.00$ & $165.$ & $0.2591$ & $  0.9324$ & $  0.2836$ & $   28.829$ & $  3.95$ & $  4.$ & $0.2997$ & $  4.4783$ & $  0.7346$ &  \\
$  3.00$ & $  0.$ & $  0.$ & $  276.224$ & $  3.00$ & $  0.$ & $0.2559$ & $  0.2885$ & $  0.1152$ & $   91.745$ & $  2.98$ & $  0.$ & $0.2730$ & $  1.6582$ & $  0.4072$ &  \\
$  3.00$ & $213.$ & $156.$ & $  338.489$ & $  3.00$ & $149.$ & $0.2580$ & $  0.6933$ & $  0.2271$ & $   77.846$ & $  2.97$ & $  5.$ & $0.3084$ & $  4.2267$ & $  0.7202$ &  \\
$  2.50$ & $  0.$ & $  0.$ & $  440.910$ & $  2.50$ & $  0.$ & $0.2559$ & $  0.2885$ & $  0.1152$ & $  175.074$ & $  2.48$ & $  0.$ & $0.2757$ & $  1.6266$ & $  0.3952$ &  \\
$  2.50$ & $204.$ & $150.$ & $  546.768$ & $  2.50$ & $143.$ & $0.2577$ & $  0.5995$ & $  0.2022$ & $  144.151$ & $  2.48$ & $  6.$ & $0.3114$ & $  3.9235$ & $  0.6792$ &  \\
\cline{10-15}
$  2.00$ & $  0.$ & $  0.$ & $  805.801$ & $  2.00$ & $  0.$ & $0.2559$ & $  0.2885$ & $  0.1152$ &  \\
$  2.00$ & $201.$ & $147.$ & $ 1017.789$ & $  2.00$ & $141.$ & $0.2579$ & $  0.5387$ & $  0.1846$ &  \\
$  1.70$ & $  0.$ & $  0.$ & $ 1284.511$ & $  1.70$ & $  0.$ & $0.2559$ & $  0.2885$ & $  0.1152$ &  \\
$  1.70$ & $188.$ & $143.$ & $ 1636.687$ & $  1.70$ & $139.$ & $0.2582$ & $  0.4852$ & $  0.1700$ &  \\
$  1.50$ & $  0.$ & $  0.$ & $ 1740.497$ & $  1.50$ & $  0.$ & $0.2559$ & $  0.2885$ & $  0.1152$ &  \\
$  1.50$ & $150.$ & $ 10.$ & $ 2076.554$ & $  1.50$ & $  9.$ & $0.2607$ & $  0.8764$ & $  0.2408$ &  \\
$  1.35$ & $  0.$ & $  0.$ & $ 2501.992$ & $  1.35$ & $  0.$ & $0.2559$ & $  0.2885$ & $  0.1152$ &  \\
$  1.35$ & $ 33.$ & $  7.$ & $ 2758.981$ & $  1.35$ & $  7.$ & $0.2600$ & $  0.5076$ & $  0.1714$ &  \\
$  1.25$ & $  0.$ & $  0.$ & $ 3298.730$ & $  1.25$ & $  0.$ & $0.2559$ & $  0.2885$ & $  0.1152$ &  \\
$  1.25$ & $100.$ & $  6.$ & $ 3438.697$ & $  1.25$ & $  5.$ & $0.2602$ & $  0.4480$ & $  0.1572$ &  \\
$  1.10$ & $  0.$ & $  0.$ & $ 4187.090$ & $  1.10$ & $  0.$ & $0.0928$ & $  0.3007$ & $  0.1114$ &  \\
$  1.10$ & $ 50.$ & $  4.$ & $ 4299.702$ & $  1.10$ & $  3.$ & $0.2477$ & $  0.2980$ & $  0.1173$ &  \\
$  1.00$ & $ 50.$ & $  3.$ & $ 6643.299$ & $  1.00$ & $  2.$ & $0.2451$ & $  0.2970$ & $  0.1169$ &  \\
$  0.90$ & $  0.$ & $  0.$ & $10021.041$ & $  0.90$ & $  0.$ & $0.1844$ & $  0.2925$ & $  0.1140$ &  \\
$  0.90$ & $ 13.$ & $  2.$ & $10591.236$ & $  0.90$ & $  1.$ & $0.2415$ & $  0.2957$ & $  0.1164$ &  \\
$  0.80$ & $  0.$ & $  0.$ & $16317.844$ & $  0.80$ & $  0.$ & $0.1842$ & $  0.2924$ & $  0.1140$ &  \\
$  0.80$ & $  7.$ & $  1.$ & $17462.567$ & $  0.80$ & $  1.$ & $0.2343$ & $  0.2924$ & $  0.1153$ &  \\
\cline{1-9}
\end{tabular}
}
\label{TabListModels}
\end{table}
\end{landscape}

\bibliographystyle{aa} 
\bibliography{GRID006} 

\end{document}